\documentclass[%
reprint,
superscriptaddress,
amsmath,amssymb,
aps,
prb,
floatfix,
]{revtex4-2}

\usepackage{graphicx}
\usepackage{dcolumn}
\usepackage{bm}
\usepackage{hyperref}
\usepackage[mathlines]{lineno}
\usepackage{amsmath,bm}
\usepackage{times}
\usepackage{fouriernc}

\begin{document}


\title{Electronic properties of topological kagome metals YV$_6$Sn$_6$ and GdV$_6$Sn$_6$}

\author{Ganesh Pokharel}
\affiliation{Materials Department and California Nanosystems Institute, University of California Santa Barbara, Santa Barbara, California 93106, USA}
\thanks{These two authors contributed equally}

\author{Samuel M. L. Teicher}
\affiliation{Materials Department and California Nanosystems Institute, University of California Santa Barbara, Santa Barbara, California 93106, USA}
\thanks{These two authors contributed equally}

\author{Brenden R. Ortiz}
\affiliation{Materials Department and California Nanosystems Institute, University of California Santa Barbara, Santa Barbara, California 93106, USA}

\author{Paul M. Sarte}
\affiliation{Materials Department and California Nanosystems Institute, University of California Santa Barbara, Santa Barbara, California 93106, USA}

\author{Guang Wu}
\affiliation{Department of Chemistry and Biochemistry, University of California Santa Barbara, Santa Barbara, California 93106, USA}

\author{Shuting Peng}
\affiliation{Hefei National Laboratory for Physical Sciences at the Microscale, Department of Physics and CAS Key Laboratory of Strongly-coupled Quantum Matter Physics, University of Science and Technology of China, Hefei, Anhui 230026, China}

\author{Junfeng He}
\affiliation{Hefei National Laboratory for Physical Sciences at the Microscale, Department of Physics and CAS Key Laboratory of Strongly-coupled Quantum Matter Physics, University of Science and Technology of China, Hefei, Anhui 230026, China}%

\author{Ram Seshadri}
\affiliation{Materials Department and California Nanosystems Institute, University of California Santa Barbara, Santa Barbara, California 93106, USA}

\author{Stephen D. Wilson}
\email{stephendwilson@ucsb.edu}
\affiliation{Materials Department and California Nanosystems Institute, University of California Santa Barbara, Santa Barbara, California 93106, USA}

\date{\today}

\begin{abstract}
The synthesis and characterization of vanadium-based kagome metals  YV$_6$Sn$_6$ and GdV$_6$Sn$_6$ are presented. X-ray diffraction, magnetization, magnetotransport, and heat capacity measurements reveal an ideal kagome network of V-ions coordinated by Sn and separated by triangular lattice planes of rare-earth ions. The onset of low-temperature magnetic order of Gd spins is detected in GdV$_6$Sn$_6$ and is suggested to be noncollinear, while V-ions in both compounds remain nonmagnetic. Density functional  theory calculations are presented modeling the band structures of both compounds, which can be classified as $\mathbb{Z}_2$ topological metals in the paramagnetic state.  Both compounds exhibit high mobility, multiband transport and present an interesting platform for controlling the interplay between magnetic order associated with the $R$-site sublattice and nontrivial band topology associated with the V-based kagome network. Our results invite future exploration of other $R$V$_6$Sn$_6$ ($R$=rare earth) variants where this interplay can be tuned via $R$-site substitution.  

\end{abstract}

\pacs{Valid PACS appear here}
\maketitle

\section{Introduction}
 
The structural motif of a kagome net of metal ions gives rise to both Dirac points in the band structure as well as destructive interference-derived flat band effects. As a result, kagome metals have the potential to host topologically nontrivial band structures intertwined with electron-electron correlation effects. Electronic instabilities resulting from this interplay have been studied theoretically ranging from bond density wave order, to charge density waves (CDW) to superconductivity \cite{PhysRevLett.110.126405,PhysRevB.87.115135,PhysRevLett.97.147202,PhysRevB.87.115135,ko2009doped,PhysRevB.86.121105,PhysRevB.80.113102, PhysRevB.104.045122}. Recent experiments have begun to probe this rich phase space and have uncovered the emergence of an unusually large anomalous Hall effect \cite{Yangeabb6003,Kida_2011,liu_sun_2018}, complex patterns of magnetism \cite{Fenner_2009, Rebacca_2021}, charge density waves \cite{Jiang2021,Zhao2021}, and superconductivity \cite{ortiz2020superconductivity, ortizCsV3Sb5, 2021Rb}, validating the promise of kagome metals to form a rich frontier of unconventional electronic phenomena.

One family of kagome metals are the so-called ``166" compounds that crystallize in the MgFe$_6$Ge$_6$ structural prototype. This class of materials is chemically very diverse, and considering the structure as $A$$B_6$$X_6$, the $A$-site can host a variety of alkali, alkali earth, and rare earth metals (e.g. Li, Mg, Yb, Sm, Gd...). The $B$-site generally hosts a transition metal (e.g. Co, Cr, Mn, V, Ni...), and the $X$-site is generally restricted to the group IV elements (Si, Ge, Sn). Due to this chemical diversity, 166 materials host a wide variety of functionalities, particularly among those with magnetic host lattices. Examples include the existence of spin polarized Dirac cones in YMn$_6$Sn$_6$ \cite{Li_2021_dirac}; large anomalous hall effects in LiMn$_6$Sn$_6$ \cite{Chen_Dong_2021}, GdMn$_6$Sn$_6$ \cite{Asaba_2020}; Chern topological magnetism in TbMn$_6$Sn$_6$ \cite{Yin2020}; competing magnetic phases in YMn$_6$Sn$_6$ \cite{Ghimire_2020}; catalytic properties in MgCo$_6$Ge$_6$ \cite{Gieck_2006}; negative magnetoreistance in YMn$_6$Sn$_{6-x}$Ga$_x$\cite{Zhang_2001}; and a cycloidal spin structure in HoMn$_{6-x}$Cr$_x$Ge$_6$ \cite{SCHOBINGER_2016}. 

One appeal of the chemical versatility of the 166 class of compounds is the ability to design materials where magnetic interactions can be tuned independently from the kagome lattice.  Nonmagnetic B-site variants, in principle, provide this flexibility and allow the interplay between magnetism and the kagome-derived band structures to be explored.  This potentially allows access to new electronic phenomena derived from coupling the triangular-lattice planes of magnetic $A$-site ions and a nonmagnetic $B$-site kagome net. Nonmagnetic kagome metals are rather underexplored relative to their magnetic counterparts, and recent investigation of nonmagnetic $A$V$_3$Sb$_5$ compounds \cite{Brenden_2019} have shown that unusual charge density wave instabilities and superconductivity may appear when local magnetic interactions are absent \cite{ortizCsV3Sb5,yuxiaoKVS}.  Finding new, nonmagnetic kagome metal variants and tuning/proximitizing magnetic coupling allowed in the 166 structure via the neighboring layers is an appealing next step in this field. 

\begin{figure*}
\centering
\includegraphics[width=1.8\columnwidth]{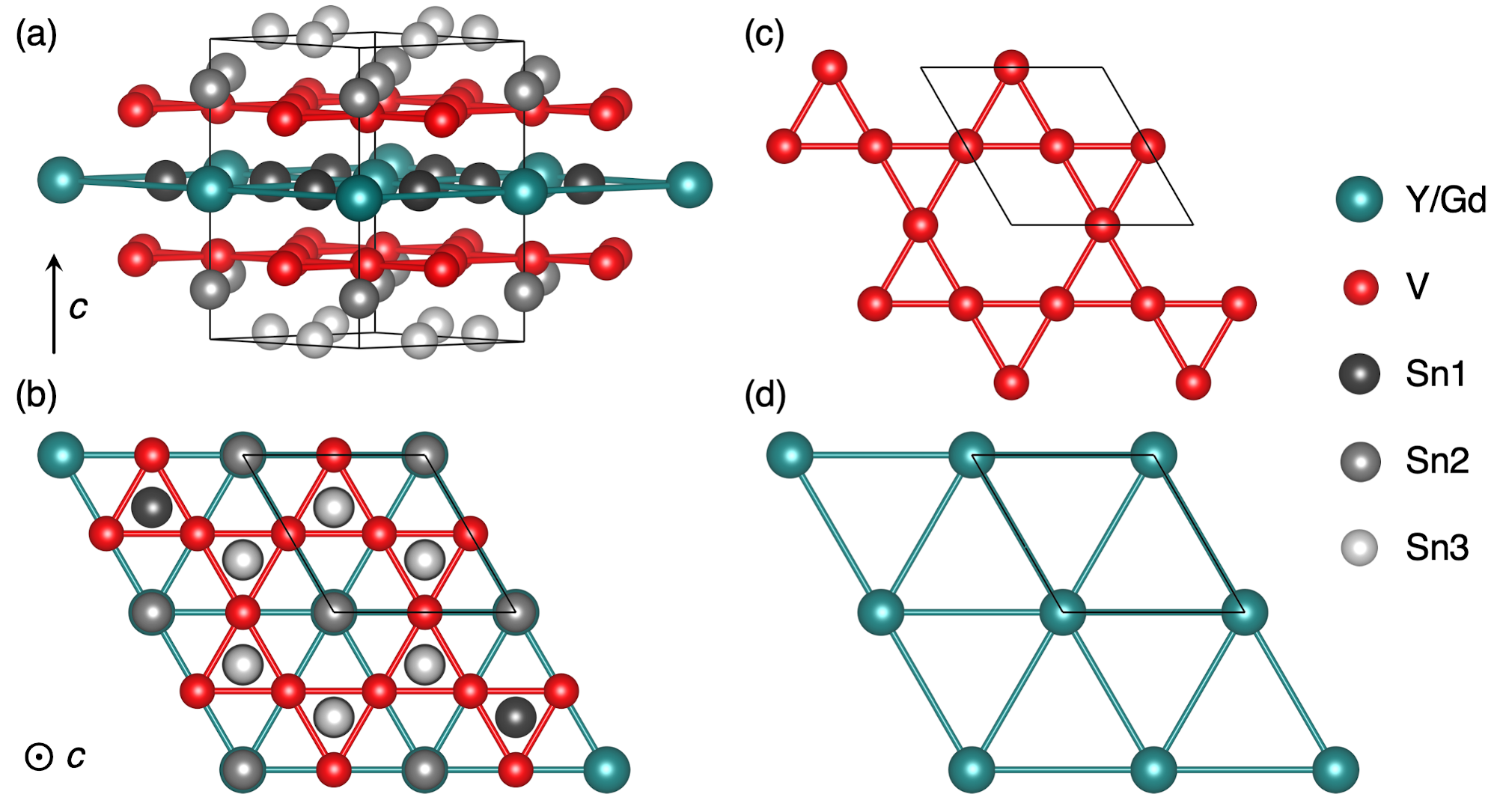}
    \caption{Crystal structure of RV$_6$Sn$_6$ (R = Gd, Y) (a) RV$_6$Sn$_6$ lattice structure comprised of different layers of V$_3$Sn2, RSn1 and Sn3 atoms. The three different types of Sn sites are represented by Sn1, Sn2 and Sn3. (b) Top view of crystal structure looking along the $c$-axis and showing the kagome plane of V-atoms and projected Sn1 and Sn3 sites. (c)  2D kagome net of V-atoms. (d) Triangular lattice of R-site (Gd, Y) ions interwoven between kagome planes as shown looking along the $c$-axis. 
    }
\label{fig1}
\end{figure*}

In this work, we report the synthesis of single crystals of YV$_6$Sn$_6$ and GdV$_6$Sn$_6$ kagome metal compounds and study their physical properties.  X-ray diffraction, magnetization, resistivity, and heat capacity data reveal multiband, high-mobility electron transport and a nonmagnetic vanadium kagome lattice.  Introduction of magnetic Gd ions results in the formation of magnetic order below 5.2 K, which, in zero field, suggests the formation of a canted or noncollinear antiferromagnetic state.  \textit{Ab initio} modeling of the band structures of these compounds establishes the presence of topological surface states at the Fermi level and categorizes the paramagnetic state as $\mathbb{Z}_2$  topological metal.  Our results demonstrate that vanadium-based 166 kagome metals are interesting platforms to studying the interplay between nontrivial band topology and correlation effects endemic to a nonmagnetic kagome lattice proximitized to magnetic order in the neighboring rare-earth layers.

\section{Experimental Details}

Single crystals of YV$_6$Sn$_6$ and GdV$_6$Sn$_6$ were synthesized via a flux-based technique. Gd (pieces, 99.9\%), Y (powder, 99.9\%), V (pieces, 99.7\%), Sn (shot, 99.99\%) were loaded inside an alumina crucible with the molar ratio of 1:6:20 and then heated at 1125$^o$C for 12 hours. Then, the mixture was cooled at a rate of 2$^o$C/h. The single crystals were separated from the flux via centrifuging at 780 $^o$C. Crystals grown via this method were generally a few millimeters in length and $<$ 1 mm in thickness. The separated single crystals were subsequently cleaned with dilute HCl to remove any flux contamination. Crystals were then transferred into a small jar of mercury to further remove additional tin contamination to the crystals.

\begin{table}
	\centering
	\caption{Structural details of YV$_6$Sn$_6$ obtained from the refinement of single crystal x-ray diffraction data at T = 300 K.  Cell refinement in $P6/mmm$ yields $\it{R_f}$= 0.0175, $\it{WR_f}$= 0.0399, and \\\    $\it{a}$ = $\it{b}$ = 5.520(2), $\it{c}$ = 9.168(4) \AA.}
	
	\begin{tabular}{c c c c c c}
		\hline
		atom (site) & $\it{x}$ & $\it{y}$ & $\it{z}$ & $\it{U_{ani}}$ & occupancy \\ [0.5ex]
		\hline\hline
		Y ($\it{1a}$) & 1.0000 & 1.0000 & 0.5000 & 0.0085(3) & 1\\ 
		V ($\it{6i}$) & 0.5000 & 0.5000 & 0.7481(1) & 0.0053(3) & 1 \\
		Sn1 ($\it{2e}$) & 1.0000 & 1.0000 & 0.8335(1) & 0.0066(2) & 1 \\
		Sn2 ($\it{2d}$) & 0.3333 & 0.6667 & 0.5000(1) & 0.0063(2) & 1 \\
		Sn3 ($\it{2c}$) & 0.3333 & 0.6667 & 1.0000 & 0.0054(2) & 1 \\ [1ex]
		\hline
	\end{tabular}
	\label{table}
\end{table}

\begin{table}
	
	\centering
	\caption{Structural details of GdV$_6$Sn$_6$ obtained from the refinement of single crystal x-ray diffraction data at T = 300 K. Cell refinement in $P6/mmm$ yields $\it{R_f}$= 0.039, $\it{WR_f}$= 0.085, and $\it{a}$ = 5.5348(7), $\it{c}$ = 9.1797(11) \AA.}
	\begin{tabular}{c c c c c c }
		\hline
		atom (site) & $\it{x}$ & $\it{y}$ & $\it{z}$ & $\it{U_{ani}}$ & occupancy \\ [0.5ex]
		\hline\hline
		Gd ($\it{1b}$) & 1.0000 & 1.0000 & 0.5000 & 0.0063(4) & 1\\ 
		V ($\it{6i}$) & 0.5000 & 0.5000 & 0.7487(2)& 0.0055(4) & 1 \\
		Sn1 ($\it{2e}$) & 1.0000 & 1.0000 & 0.8344(1) & 0.0072(4) & 1 \\
		Sn2 ($\it{2d}$) & 0.3333 & 0.6667 & 0.5000 & 0.0054(4) & 1 \\
		Sn3 ($\it{2c}$) & 0.3333 & 0.6667 & 1.0000 & 0.0064(4) & 1 \\ [1ex]
		\hline
	\end{tabular}
\end{table}

Single-crystal x-ray diffraction measurement were carried out on a Kappa Apex II single-crystal diffractometer with a charge coupled device (CCD) detector and a Mo source. Structural solutions were obtained using the SHELX software package \cite{Shelx_2015}. Powder x-ray diffraction (PXRD) measurements were performed on a Panalytical Empyrean powder diffractometer using powdered single crystals. This was done to further verify the structure and phase purity over a larger volume. 

Magnetization measurements were carried out using a Quantum Design Magnetic Properties Measurement Systems (MPMS-3). Plate-like single crystals were attached to quartz paddles using GE-Varnish. Measurements were carried out with the magnetic field applied parallel to and perpendicular to the c-axis from 2 K to 300 K. The crystals were polished gently on the top and bottom surfaces prior to measurement. For heat capacity measurements,  crystals of mass 3.68 mg (YV$_6$Sn$_6$) and 1.69 mg (GdV$_6$Sn$_6$) were mounted to the addenda using N-grease.  Longitudinal, transverse and Hall magnetoresistance measurements were carried out using the electrical transport option (ETO) of the Quantum Design Dynacool Physical Properties Measurement System. Four-point measurements were used.

\section{Computational Methods}

\textit{Ab initio} simulations were completed in VASP \cite{Kresse1994,Kresse1996a,Kresse1996b} using the PBE functional \cite{Perdew1996} with projector-augmented waves, \cite{Blochl1994a,Kresse1999}. PAW potentials for V and Sn were selected based on the VASP $v$5.2 recommendations. For the calculations presented in the main text, Gd potentials with a frozen $f$-orbital core were chosen in order to approximate the paramagnetic phase previously investigated in ARPES experiments \cite{Peng2021}. In the supporting material, electronic structure calculations are completed for the low-temperature ferromagnetic phase using complete Gd potentials with a Hubbard potential $U=$ 6\,eV applied to the Gd $f$ orbitals. This choice of $U$ gives a magnetic moment $\mu\approx$ 7\,$\mu$B, consistent with experiment (a Hubbard $U$ correction near 6\,eV is generally expected for Gd \cite{Anisimov1997}). Calculations employed an 11$\times$11$\times$5 $\Gamma$-centered $k$-mesh and a plane wave energy cutoff of 400\,eV. Structures were relaxed in VASP \textit{via} the conjugate gradient descent algorithm with a force-energy cutoff of $10^{-4}$\,eV. All calculations after relaxation employed spin-orbit coupling corrections with an energy convergence cutoff of $10^{-6}$\,eV. Tight-binding models were constructed by projecting onto valence orbitals (Gd d; V d; Sn p; inner window $E_F \pm 2$\,eV; outer window $E>E_F-5.3$\,eV) using the disentanglement procedure in \textsc{Wannier90} \cite{Mostofi2014}. Surface state Green's function calculations were completed in the \textsc{Wannier Tools} package \cite{Wu2018,Sancho1985}. Irreducible representations used to determine the $\mathbb{Z}_2$ invariant were determined with \textsc{Irvsp} \cite{Gao2021}. COHP calculations and orbital projections employed \textsc{LOBSTER}; these calculations do not incorporate spin-orbit coupling, which is not implemented in \textsc{LOBSTER} \cite{Dronskowski1993,Deringer2011,Maintz2013,Maintz2016}. A Gaussian smoothing with standard deviation 0.1\,eV was applied to the density of states and COHPs. Structures were visualized with \textsc{VESTA} \cite{Momma2011}. Additional computational details, including an initial comparison of the \textit{ab initio} band structure to experimental ARPES measurements, a comparison of the relaxed vs. experimental lattice parameters and the full $\mathbb{Z}_2$ invariant calculations are available in the supporting material.

\section{Results and Discussion} 

\subsection{Crystal structure}

\begin{figure}
\centering
\includegraphics[width=1\columnwidth]{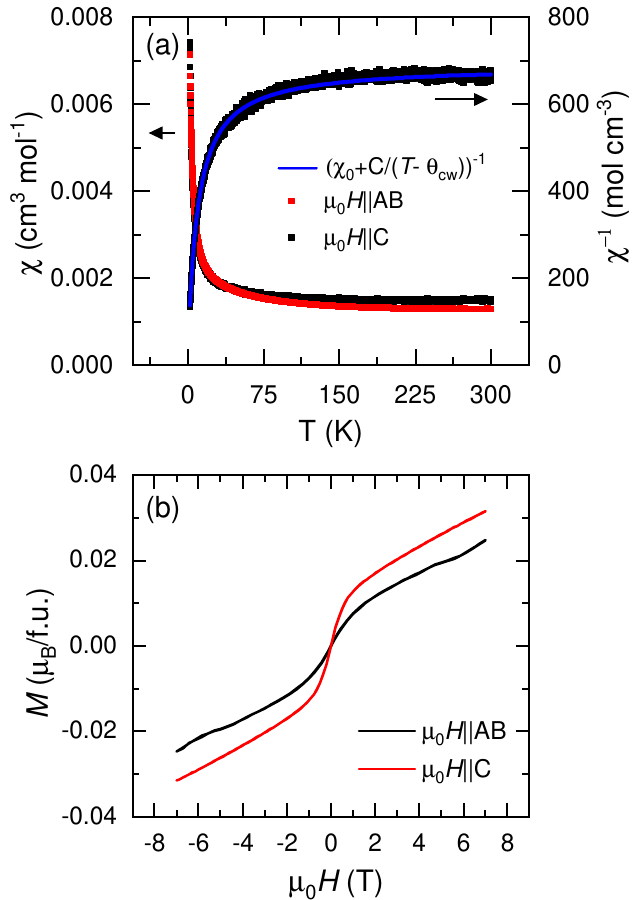}
    \caption{Magnetization measurements from YV$_6$Sn$_6$. (a) Temperature-dependent magnetic data plotted as magnetic susceptibility, $\chi=M/H$, and inverse magnetic susceptibility, $\chi^{-1}$, collected with 100 mT applied parallel and perpendicular to the $c$-axis. Horizontal arrows indicate the corresponding axis for the data. The blue curve shows the result from a Curie-Weiss fit to the data with a temperature independent $\chi_0$ term. (b) Field-dependent magnetization collected at 2 K with the field applied parallel and perpendicular to the $c$-axis. The non-linear field dependence arises due to quantum oscillations and the de Haas-van Alphen effect. The rapid upturns at low-T and low-H are due to the presence of weak paramagnetic impurities in the sample.
    }
\label{fig2}
\end{figure}

\begin{figure}
\centering
\includegraphics[width=1\columnwidth]{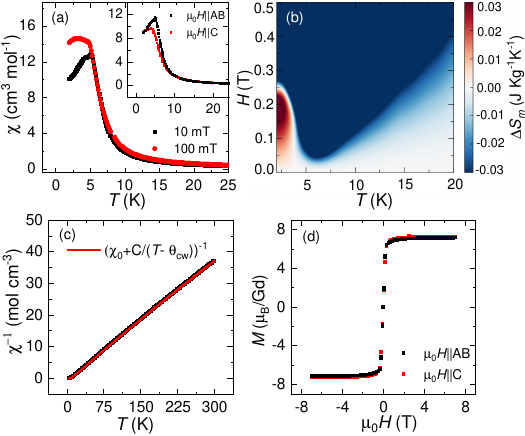}
    \caption{Magnetization measurements from GdV$_6$Sn$_6$. (a) Temperature-dependent magnetization data plotted as magnetic susceptibility, $\chi=M/H$, at fields of $\mu_0$H = 10  and 100 mT applied perpendicular to the $c$-axis. A magnetic transition is observed at T$_m$ $\sim$ 5.2 K. The inset shows a comparison of $\chi$ for a magnetic field applied parallel and perpendicular to the $c$-axis. Weak anisotropy is observed near T$_m$. (b) Magnetoentropy map ($\Delta S_m (T, H)$) of GdV$_6$Sn$_6$ near the ordering temperature for the fields applied perpendicular to the $c$-axis. (c) Temperature-dependent inverse susceptibility, $\chi^{-1}$, for fields applied parallel to the $c$-axis. The red solid line shows a Curie-Weiss fit to the data as described in the text. (d) Field-dependent magnetization data collected at 2 K for the field applied both parallel and perpendicular to the $c$-axis.
    }
\label{fig3}
\end{figure}

\begin{figure}
\centering
\includegraphics[width=0.8\columnwidth]{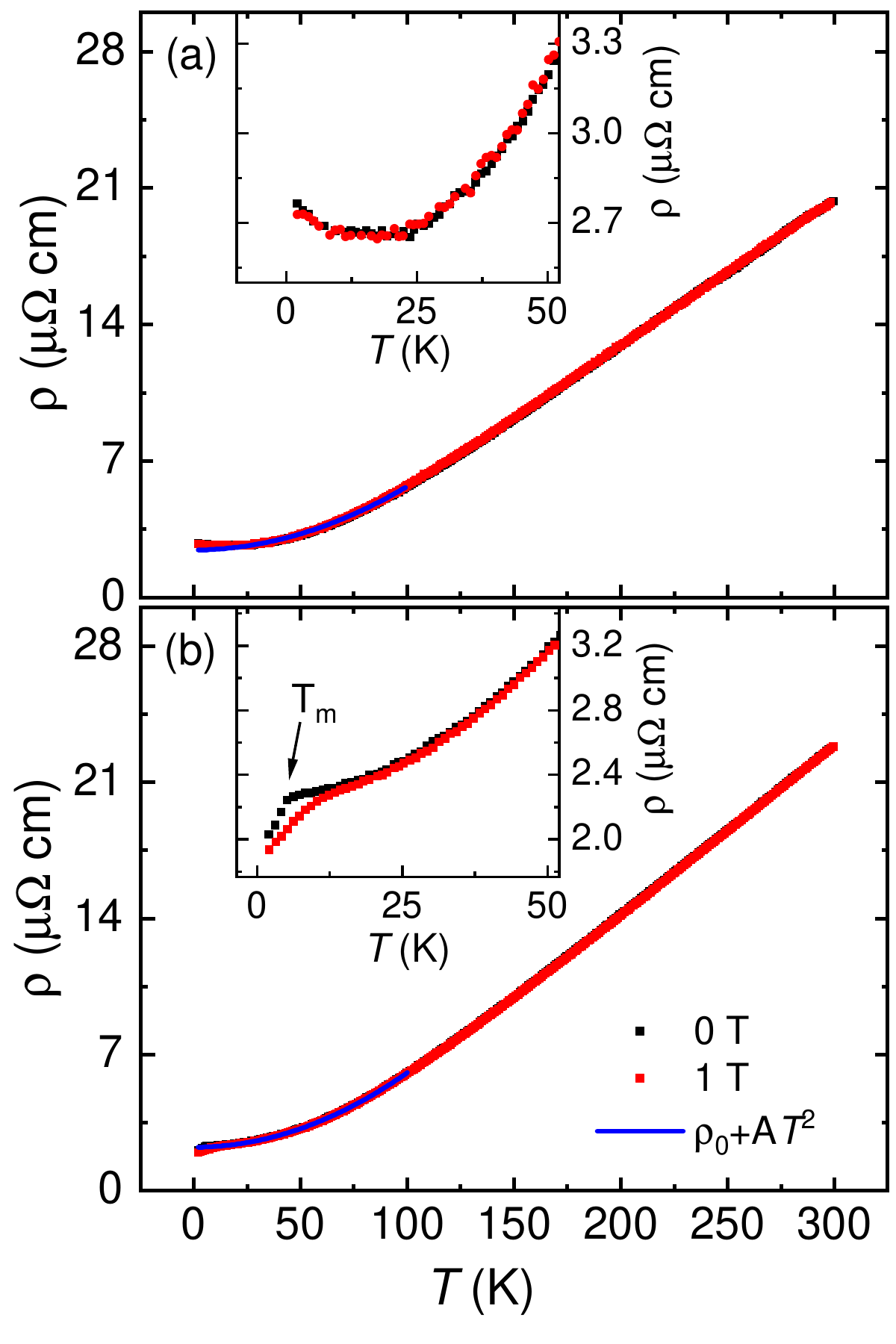}
        \caption{Electrical transport measurements collected from YV$_6$Sn$_6$ and GdV$_6$Sn$_6$ with current flowing in the $ab$-plane. $\rho(T)$ is plotted for YV$_6$Sn$_6$ and GdV$_6$Sn$_6$ in panels (a) and (b) respectively.  Data were collected under both 0 T and with 1 T applied along the $c$-axis. The blue curve shows an effective parameterization of the data via equation $\rho = \rho(0) +AT^2$.
    }
\label{fig4}
\end{figure}

The crystal structure of RV$_6$Sn$_6$ (R = Y, Gd) was obtained from the refinement of x-ray single crystal diffraction data and the structure is illustrated in Fig. \ref{fig1}. YV$_6$Sn$_6$ and GdV$_6$Sn$_6$ both exhibit the MgFe$_6$Ge$_6$-type structure with a stacking of the kagome layers of V-ions along the cryostallographic c-axis.
 The Y/Gd ions as well as the vanadium ions occupy unique crystallographic sites; whereas Sn ions occupy three different types of crystallographic sites denoted by Sn1, Sn2 and Sn3 in Fig. \ref{fig1}. A unit cell consists of the layers of V$_3$Sn2 separated by two inequivalent layers of Sn3 and RSn1, forming [V$_3$Sn2][RSn1][V$_3$Sn2][Sn3] layers along the c-axis. Fig. \ref{fig1}(b) reveals the topside view of the crystal structure where the V-atoms form a kagome layer within the ab-plane. Sn2 and Sn3 sites form stannene planes between the kagome layers of V atoms. The isolated kagome net of V atoms is shown in Fig. \ref{fig1}(c). The interstitial rare-earth atoms form a triangular lattice plane as shown in Fig. \ref{fig1}(d). 
 
 The refined structural parameters of  YV$_6$Sn$_6$ and GdV$_6$Sn$_6$ are shown in Table \ref{table}.  Nearest neighbor distances within the kagome plane are reasonably close with V-V distances being 2.76 \AA {} in YV$_6$Sn$_6$ and 2.77 \AA{} in GdV$_6$Sn$_6$. Sn2 atoms center laterally within the hexagons of the V-based kagome plane and are displaced slightly upward/downward along the c-axis. This is analogous to the CoSn-B35 type structure where the R sites are empty and the Sn atoms reside within the kagome planes of Co-atoms \cite{LARSSON199679}. In RV$_6$Sn$_6$, steric effects introduced by the R atoms push the Sn2 atoms out of the kagome layer, and this arrangement is distinct from the structures of other well-known Sn-based kagome metals such as Fe$_3$Sn$_2$ \cite{Malaman_1976, Lin_2020} and Co$_3$Sn$_2$S$_2$ \cite{Guin_2019} where the Sn atoms almost lie within the kagome layers of Fe and Co atoms respectively. 
 
\subsection{Magnetic properties}

Temperature-dependent magnetization measurements were carried out with the magnetic field applied both parallel and perpendicular to the (001) crystal surface. The results are summarized in Figures \ref{fig2} and \ref{fig3}. In  YV$_6$Sn$_6$ (Fig. \ref{fig2}), the measured magnetic susceptibility is small ($\approx$ 10$^{-3}$ cm$^3$ mol$^{-1}$) and, at high temperature, it shows predominantly Pauli-like, paramagnetic behavior. A weak upturn is observed upon cooling, likely due to a small fraction of impurity moments. To quantify this, the composite $\chi$(T) is fit to a modified Curie-Weiss form adding a $\chi$(0) contribution. This yields a $\chi$(0) of 0.00145(6) cm$^3$ mol$^{-1}$,  Curie constant of 0.01319(3) cm$^3$ mol$^{-1}$K$^{-2}$ and $\Theta_{CW}$ of -0.3(2) K. The Curie constant results in a weak effective moment of 0.3(1) $\mu_B$/f.u and $\Theta_{CW}$ is zero within the error of the fit. The small anisotropy in $\chi$ with the field parallel and perpendicular to the $c$-axis is small and likely extrinsic.  It probably arises from massing error associated with the measurement of two different crystals combined with small, uncorrected, demagnetization factors. 

Fig. \ref{fig2}(b) shows isothermal magnetization data from YV$_6$Sn$_6$ collected at 2 K. The low-field magnetization is dominated by the rapid polarization of the small impurity fraction, which at high magnetic field evolves into linear behavior from the dominant, Pauli-like susceptibility. Oscillations become apparent in the high-field magnetization when the field is aligned within the ab-plane due to the de Haas-van Alphen effect, suggesting a reasonably high mobility for the charge carriers.  This is explored in greater depth in the next section.

Turning to GdV$_6$Sn$_6$, an anomaly in the magnetic susceptibility, $M/H=\chi$, is evident at T$_m= 5.2(2)$ K, indicating a magnetic transition (Fig. \ref{fig3}). On cooling, a rapid divergence in the magnetization suggests the onset of ferromagnetic correlations; however the moment decreases below T$_m$ under low field ($H=10$ mT) measurements. The low-temperature downturn in magnetization persists in both zero-field and field-cooled measurements (not shown), suggesting antiferromagnetic correlations rather than conventional glass-like moment freezing.  Upon applying a slightly higher magnetic field ($H=100$ mT), the low-temperature downturn is largely suppressed and a ferromagnetic state is polarized as shown in the Fig. \ref{fig3}(a).  The inset of Fig. \ref{fig3}(a) shows a comparison of $\chi$ for the magnetic field applied perpendicular and parallel to the c-axis. A weak anisotropy is observed below T$_m$ suggesting a slight easy-plane anisotropy. The field and temperature dependence of the field-polarized ferromagnetic state, plotted as a magnetoentropy map $\Delta S_m (T, H)$ \cite{Bocarsly2018}, is shown in Fig. \ref{fig3}(b).  The $\Delta S_m (T, H)$ map indicates that the low-temperature phase boundary between the field-induced ferromagnetic state and the low field, likely non-collinear, magnetic state is near 200 mT.

Fig. \ref{fig3}(c) shows the inverse susceptibility, $1/\chi$, collected under 10 mT with a crystal of mass 0.26 mg. $1/\chi$ is linear above the magnetic ordering temperature and Curie-Weiss analysis incorporating a small $\chi_0$ term of the form $\chi= \chi(0) + C/(T-\Theta_{CW})$ was performed above 100 K, yielding $C=8.28(1)$ cm$^3$ K mol$^{-1}$, $\Theta_{CW}=7.56(2)$ K for the field applied perpendicular to the $c$-axis and $C=7.33(1)$ cm$^3$ K mol$^{-1}$, $\Theta_{CW}=7.76(2)$ K for the field applied parallel to the c-axis. A positive $\Theta_{CW}$ is consistent with the presence of ferromagnetic correlations above  T$_m$, and the effective paramagnetic moment $\mu_{eff}=7.90(3)$ $\mu_B$/Gd is consistent with the that expected for $J=7/2$ Gd$^{3+}$ spins and the low-lying crystal field multiplets populated above 100 K. Isothermal magnetization data at 2 K are plotted in Fig \ref{fig3}(d). The magnetization rapidly increases with applied field and reaches saturation slightly below $\mu_0 H=1$ T.   The  moment saturates near the expected value of 7 $\mu_B$/Gd-ion for both field parallel and perpendicular to the crystal surface, and no hysteresis is observed.

\begin{figure}
\centering
\includegraphics[width=1\columnwidth]{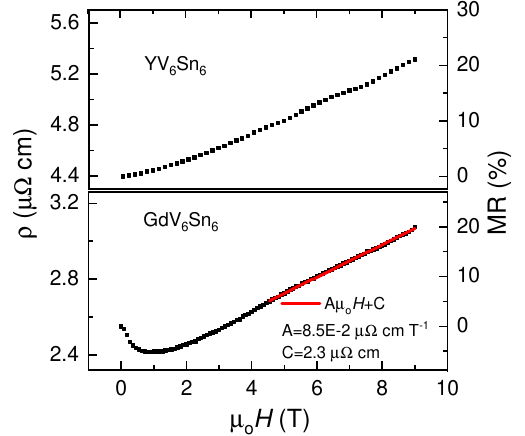}
    \caption{Transverse magnetoresistance data collected from GdV$_6$Sn$_6$ (a) and YV$_6$Sn$_6$ (b). Isothermal $\rho$(H) data exhibited weak asymmetry upon reversed field direction due to the contamination from Hall signal. The magnetoresistance component was isolated by averaging $\rho_{xx}(H)$ and $\rho_{xx} (-H)$. At high fields, $\rho$(H) displays linear field dependence in GdV$_6$Sn$_6$ whereas a weak oscillating behavior is resolved in YV$_6$Sn$_6$. The solid red line in (a) represents a linear fit to the high field $\rho$(H) of GdV$_6$Sn$_6$.  } 
    
\label{fig5}
\end{figure}

\subsection{Transport properties}
\label{Trp}
Electrical resistivity data $\rho_{xx} (T,H)$ were collected as a function of temperature and magnetic field for both YV$_6$Sn$_6$ and GdV$_6$Sn$_6$. Temperature dependent $\rho_{xx} (T,H)$ measurements at both 0 T and 1 T are shown for YV$_6$Sn$_6$ and GdV$_6$Sn$_6$ in Figs. \ref{fig4}(a) and \ref{fig4}(b) respectively. Crystals show a residual resistivity ratio $\rho_{xx}(300$ K$)/\rho_{xx}(2$ K$)\approx10$.  YV$_6$Sn$_6$ exhibits a weak upturn below $\approx 15$ K, while GdV$_6$Sn$_6$ shows an inflection in $\rho_{xx}(T)$ when cooling below $T_m$.  Above these low-temperature anomalies, $\rho(T)$ largely follows conventional Fermi-liquid $\rho\propto T^2$ behavior. 

The low-temperature resistivity data in the paramagnetic state were fit to the form $\rho_{xx} (T)$ = $\rho $(0) + A$T^2$ below 100 K.  The resulting fits are shown as blue curves in Figs. \ref{fig4}(a) and \ref{fig4}(b) with $\rho_{xx}(0)=2.40(7)$ $\mu\Omega$ cm and A=0.00032(1) $\mu\Omega$ cm K$^{-2}$ for YV$_6$Sn$_6$ and $\rho_{xx}(0)= 2.20(4)$ $\mu\Omega$ cm and A= 0.00038(1) $\mu\Omega$ cm K$^{-2}$ for GdV$_6$Sn$_6$.  At high temperature, $\rho_{xx}(T)$ evolves into a conventional, linear form.

\begin{figure*}
\centering
\includegraphics[width=2\columnwidth]{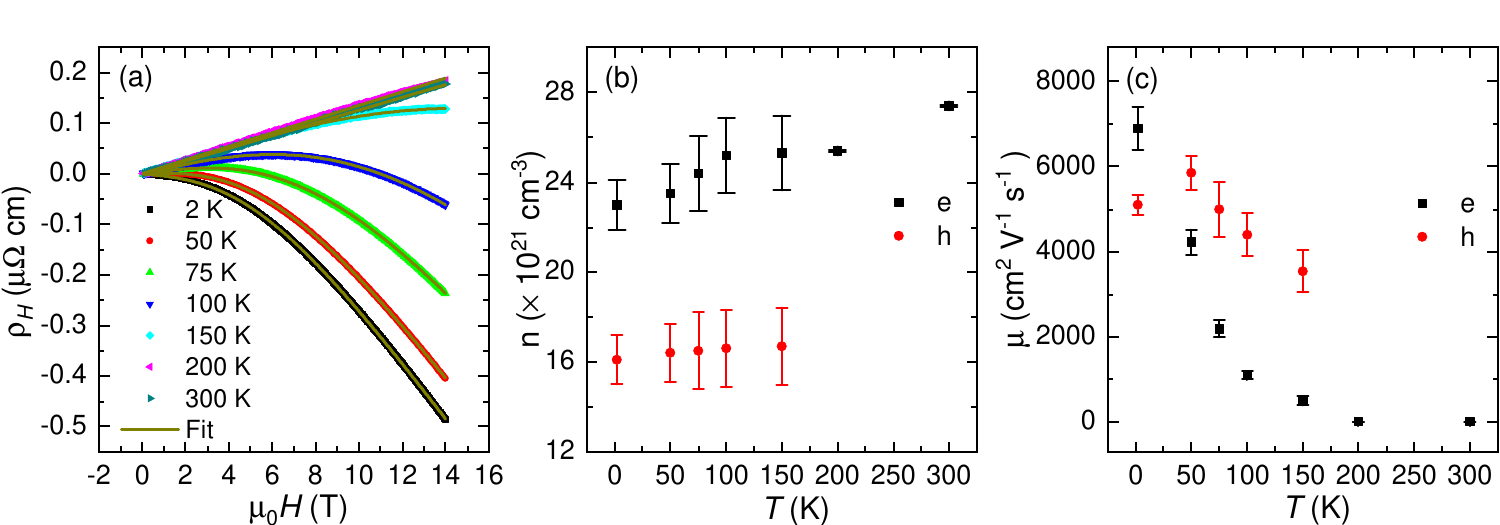}
    \caption{Hall effect measurements on YV$_6$Sn$_6$. (a) Field dependence of the Hall resistivity $\rho_{xy}$, measured at different temperatures.  To remove contributions from the longitudinal resistivity, $\rho_{xy}$ is isolated via $\rho_{xy} =  (\rho_{xy}(H)-\rho_{xy}(-H)$)/2. Solid lines through the data show fits to a two-band model below 150 K and single band model above 200 K. (b) Temperature dependence of electron-type and hole-type carrier densities, $n$, obtained by the fits in panel (a). Electron and hole-type charge carriers are denoted by e and h in the figure. (c) Temperature dependence of the charge carrier mobilities, $\mu$ for each carrier type obtained from fits to the Hall data.
    }
\label{fig6}
\end{figure*}

The isothermal, transverse magnetoresistance $\rho(H)$ and the resulting magnetoresistance ratio (MR) at 2 K are plotted in Figs. \ref{fig5}(a) and \ref{fig5}(b) for YV$_6$Sn$_6$ and GdV$_6$Sn$_6$ respectively. Nonsaturating, positive MR is observed in both compounds up to 9 T. $\rho(H)$ shows a weak asymmetry when the direction of applied field is reversed, which arises due to the mixing of the magnetoresistance with a strong Hall signal. This mixing is removed and the MR isolated by examining the symmetric component via averaging the $\rho(H)$ values over the positive and negative field directions \cite{Feng_2015, Visser_2006}. The even component is plotted as $\rho(H)$ in Figs. \ref{fig5}(a) and \ref{fig5}(b). 

At low fields, YV$_6$Sn$_6$ shows a weakly quadratic MR, that evolves into a quasi-linear MR in the high field limit.  Weak oscillations are dressed on top of a linear MR, reflective of Shubnikov de Haas quantum oscillations in the magnetotransport, and are consistent with the de Haas van Alphen quantum oscillations observed in magnetization data.  More in-depth analysis of the quantum oscillations will require higher field strengths and lower temperatures; however Hall measurements can be used to further characterize the mobilities and carrier concentrations of the conduction bands. In contrast, magnetic GdV$_6$Sn$_6$ shows a negative MR at low-fields, which reaches a minimum near 1 T as the system enters the saturated, ferromagnetic state.  Upon increasing field within the magnetically polarized state, the MR switches sign to become positive and linear at the high-field limit.

To better characterize the nature of conduction in YV$_6$Sn$_6$, Hall $\rho_{H}(T,H)$ measurements were performed. Fig. \ref{fig6}(a) shows $\rho_H(H)$ measured at various temperatures with the magnetic field aligned parallel to the c-axis. At 2 K, $\rho_H$ is not linear with field at low temperature, signifying the presence of multi-band transport. As the temperature is increased, a linear, single-band Hall response appears---by 300 K the transport can be described via an effective single-carrier model.

 To parameterize the Hall data, $\rho_H$ is fit with two different models at low ($T < 150$ K) and high ($T > 150$ K) temperature, where two-band and single-band fits, respectively, best describe the data. Based on the two band model, Hall resistivity is fitted with equation

\begin{equation}
\rho_H = \frac{H}{e} \frac{(n_h \mu_h^2-n_e\mu_e^2)+ (n_h-n_e)\mu_h^2\mu_e^2H^2}{(n_h  \mu_h+n_e\mu_e)^2+(n_h-n_e)^2\mu_h^2\mu_e^2H^2} 
\label{eqn}
\end{equation} 

Where, $n_h$, $n_e$, $\mu_h$, $\mu_e$ represents the carrier density and mobility of holes and electrons respectively.

 Fits are shown as lines in Fig. \ref{fig6}(a). Above 150 K, the $\rho_H$ data are dominated by electron-like carriers whose mobility increases quickly upon cooling.  Below 150 K, hole-like carriers contribute to the measured Hall response and the mobilities of each carrier type become comparable at the lowest measured temperature, 2 K.  The carrier density, $n$, and mobility, $\mu$, obtained from the fits are plotted in Fig \ref{fig6}(b) and \ref{fig6}(c) respectively.
 
 
\subsection{Heat capacity measurements}

\begin{figure}
\includegraphics[width=\columnwidth]{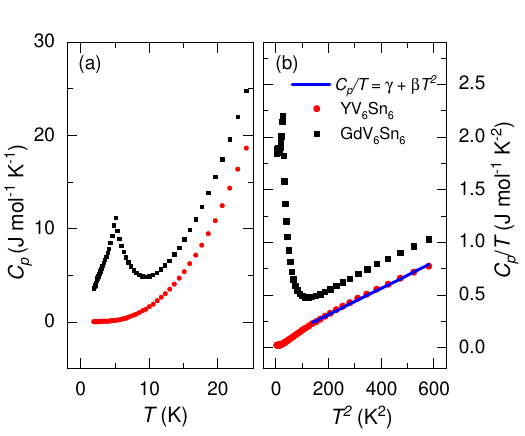}
    \caption{(a) Temperature dependent heat capacity, $C_p(T)$, of GdV$_6$Sn$_6$ and YV$_6$Sn$_6$ measured in zero-magnetic field. (b) Temperature dependence of $C_p/T$ plotted versus $T^2$. To calculate the electronic and phonon contributions,  Solid line shows the fit to the form  $C_p(T)=\gamma T$ + $\beta T^3$ as described in the text.}
\label{fig7}
\end{figure}

Temperature-dependent heat capacity, $C_p(T)$, data characterizing both YV$_6$Sn$_6$ and GdV$_6$Sn$_6$ were collected and are summarized in Fig. \ref{fig7}. $C_p$ measurements reveal features consistent with both transport and magnetization data. Specifically, anomalies appear in the low temperature $C_p$ for both compounds. In GdV$_6$Sn$_6$ a sharp cusp is observed at 5 K, coinciding with the onset of magnetic order below $T_m$. In the case of YV$_6$Sn$_6$, $C_p$ follows the form $C_p$ = $\gamma T$ + $\beta T ^3$ at low temperature; however fits to this form break down below $\approx 12$ K.  This breakdown occurs near the minimum in resistivity data shown in Fig. 4 (b), suggesting that the apparent localization of carriers in charge transport coincides with the depletion of the density of states sampled in $C_p$ data.  Above this breakdown, a fit parameterizing the Sommerfeld coefficient in YV$_6$Sn$_6$ is shown in Fig. 7 (b) with the effective coefficient prior to the suppression in the density of states being $\gamma=0.067(6)$ J mol$^{-1}$ K$^{-2}$. 

\begin{figure}
	\includegraphics[width=0.4\textwidth]{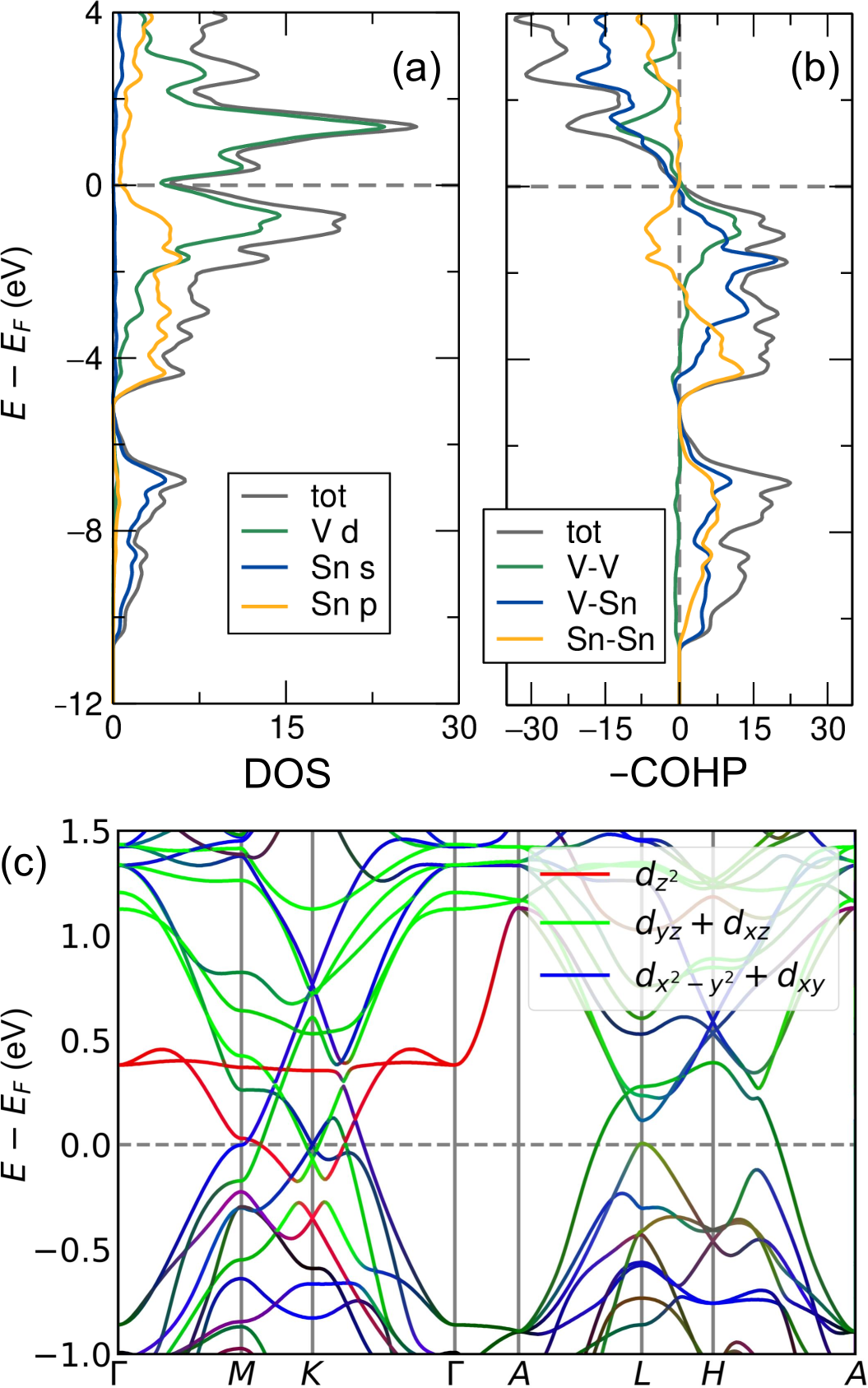}
	\caption{\label{fig:Figure2} Orbital origins of electronic structure. (a) orbital-projected density of states showing that the electronic structure near $E_F$ derives primarily from the V $d$ states. (b) crystal orbital Hamilton population curves for V-V, V-Sn, and Sn-Sn bonding.   (c) V $d$ orbital decomposed band structure.}
\end{figure}

\section{Electronic structure}

\begin{figure}
	\includegraphics[width=0.4\textwidth]{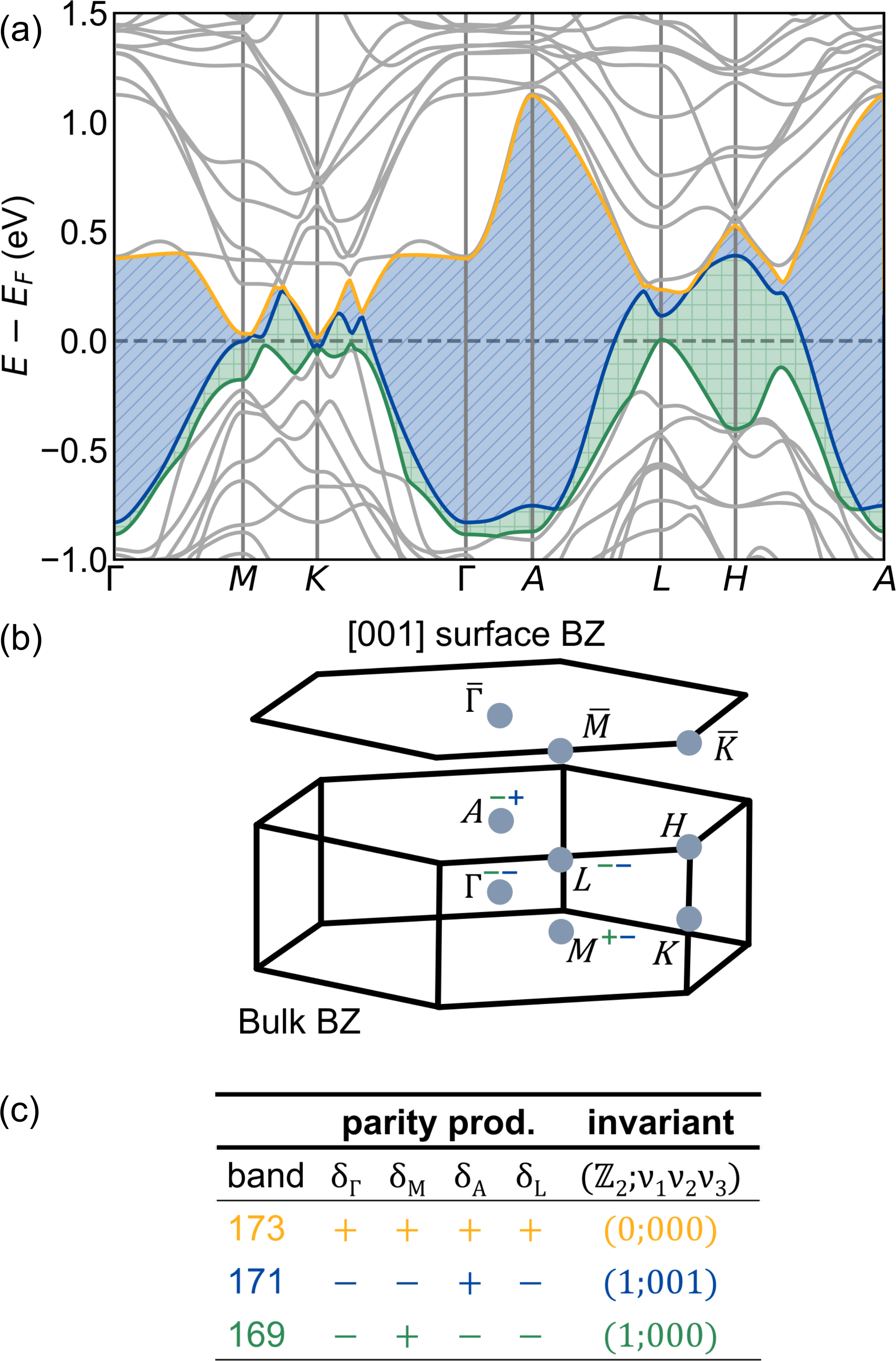}
	\caption{\label{fig:Figure3} Topological classification. (a) band structure with bands 169 (green), 171 (blue), and 173 (orange). Shaded regions show ($E$, $k$) space where topologically nontrivial states are expected to appear (b) visualization of the hexagonal Brillouin zone and projected [001] surface Brillouin zone, identifying high symmetry points. (c) Parity products classifying the $\mathbb{Z}_2$ invariant for each band. Bands 169 and 171 are characterized by a strong topological invariant, $\mathbb{Z}_2=1$. Band 173 is trivial with no topological invariants. In addition to the strong invariant, band 171 also supports a weak invariant $\nu_3=1$.}
\end{figure}

The electronic structure of (Y,Gd)V$_6$Sn$_6$ was modeled via density functional theory calculations. YV$_6$Sn$_6$ and GdV$_6$Sn$_6$ show qualitatively similar band structures in the paramagnetic state and for clarity, we focus on the electronic structure of GdV$_6$Sn$_6$ in the paramagnetic phase in the following paragraphs.  

Figure 8 shows the orbitally decomposed electronic structure of GdV$_6$Sn$_6$ with the orbital breakdown of the density of states shown in Fig. 8 (a). Fig. 8 (b) shows the crystal orbital Hamilton population curves projected for V-V, V-Sn, and Sn-Sn bonding interactions, where all are shown to contribute significantly near the Fermi-level. V-Sn and V-V bands are approximately half-filled, whereas states arising from the Sn $p$-Sn $p$ interaction are fully filled. As a result, filled Sn $p$-Sn $p$ antibonding states contribute near the Fermi level and likely play an important role in the structure.  Fig. 8 (c) shows the V-based $d$-orbital band structure endemic to the kagome lattice. A prominent $d_z^2$ kagome flat-band can be seen above $E_F$, and Dirac cones and saddle points similar to those expected from minimal kagome tight-binding models lie at the Fermi level. Given the local kagome coordination in this structure, it is unsurprising that out-of-plane $d_z^2$ orbital states fill prior to $d_{yz}$ + $d_{xz}$ and in-plane $d_{x^2-y^2}$ + $d_{xy}$ states.

Figure 9 shows the band structure of GdV$_6$Sn$_6$ in the paramagnetic phase alongside the topological classification of the metallic state based on the bands crossing $E_F$.  The band structure agrees well with ARPES data measuring the band structure \cite{ESI}. Bands crossing $E_F$ are highlighted in Fig. 9 (a) with high symmetry points labeled in Fig. 9 (b) for reference.  Due to the presence of small, but continuous gaps between bands, the $\mathbb{Z}_2$ topological classification can be determined for each band using parity products, and a strong topological invariant $\mathbb{Z}_2=1$ can be assigned to bands 171 (blue) and 169 (green), while the topmost band 173 (yellow) is topologically trivial. As a result of these invariants, topological surface states are expected in the gaps between bands 169 and 171 (green, square-hatched) as well as between bands 171 and 173 (blue, diagonal-hatched) marked in Fig. 9 (a). Further classification is presented in the supplemental material accompanying this paper \cite{ESI}.

Exploring the possibility of topologically nontrivial surface states further, Fig. 10 plots projections of predicted surface states along the [001] surface with a Gd-terminated surface in Fig. 10 (a) and a Sn3-terminated surface in Fig. 10 (b).  Comparing the two plots, many bright surface state bands can be identified in (b) which are not present in the bulk. Near-Fermi level surface bands can be seen emitting from the bulk Dirac cones on either side of $\overline{K}$. A pair of surface states bridge the large local band gap at $\overline{\Gamma}$, with a surface Weyl band crossing appearing at $E\approx-0.4$\,eV. The presence of this rich surface state spectrum is expected from the topological invariant calculation described in Figure \ref{fig:Figure3}.

\begin{figure}
	\includegraphics[width=0.4\textwidth]{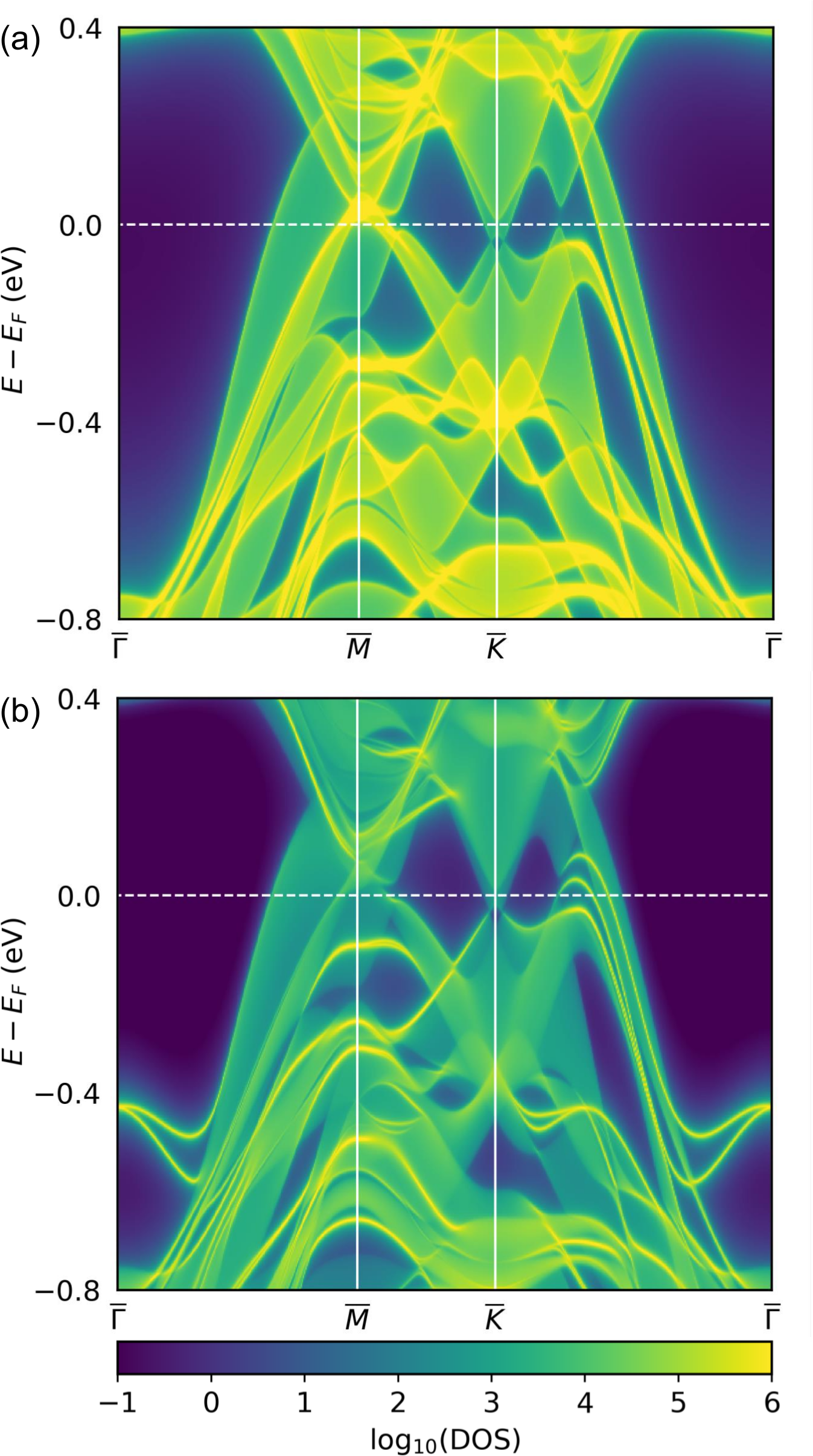}
	\caption{\label{fig:Figure4} [001] surface states. (a) and (b) display the surface Green's function projection of pure bulk states and the states on a Gd/Sn$_1$ terminated surface, respectively. }
\end{figure}

\section{Discussion}

The nature of the metallic state associated with the V-based kagome net can be probed by examining the electronic properties of the nonmagnetic YV$_6$Sn$_6$ compound. Using the Sommerfeld coefficient $\gamma$, the effective quadratic coefficient $A$ of the temperature dependent resistivity $\rho(T)$, and $\chi_0$ from susceptibility data, comparative metrics can be established. Using these, the Sommerfeld-Wilson ratio, proportional to the ratio of Pauli susceptibility to the Sommerfeld coefficient, is $R=2.23$ in units of 3$\mu_B^2/4\pi^2K_B^2$ and is slightly enhanced above the nominal $R=1$ for an uncorrelated metal. The Kadowaki-Woods ratio ($A/\gamma^2$) \cite{KADOWAKI_1986} similarly provides a slightly enhanced value of $7.1\times10^{-2}$ $\mu\Omega$ cm mol$^2$ K$^2$ J$^{-2}$ for YV$_6$Sn$_6$. These values suggest moderate correlation effects may be relevant for electron transport in these compounds; however we caution that the $A$ and $\gamma$ values are only effective parameters extracted prior to the low-temperature upturn in resistivity and coincident suppression in the low-temperature density of states of this system. 

Low-temperature Hall effect data show a multiband character for the electron transport onsets upon cooling.  This behavior can be parameterized via two-band fits, and at low temperatures, both electron-like and hole-like bands develop reasonably high mobilities of nearly 6000 cm$^2$ V$^{-1}$ s$^{-1}$. At 2 K, this results in the onset of weak, quantum oscillations in both the magnetoresistance as well as the magnetization.  Future measurements at higher fields and lower temperatures will be required to fully map these oscillations and connect them to extremal orbits modeled via \textit{ab initio} models of the band structure.

Characterizing similar properties in the GdV$_6$Sn$_6$ compound is complicated by the presence of magnetic order and the influence of spin correlations in the low-temperature electron transport and heat capacity.  The onset of ferromagnetic correlations is interrupted by an apparent freezing below $T_m=5$ K and a potential second transition apparent as an inflection at 3 K in the ZFC data.  Both features are rapidly quenched with increasing magnetic field.  This and the absence of a low-field hysteresis in the magnetization data suggest that a weak, noncollinear magnetic order forms below 5 K, though future magnetic scattering (neutron or resonant x-ray) measurements are required to verify this.

Similar to the recently reported AV$_3$Sb$_5$ compounds \cite{ortiz2020superconductivity, ortizCsV3Sb5}, the band structures of both YV$_6$Sn$_6$ and GdV$_6$Sn$_6$ in the paramagnetic phase can be categorized as $\mathbb{Z}_2$ topological metals with surface states predicted at E$_F$.  Furthermore, a clear flat band appears in the band structure $\approx 400$ meV above E$_F$, consistent with the interference effects expected from a kagome-derived band. Multiple Dirac points appear near E$_F$ at the $K$ points as well as a van Hove singularity (vHs) near the M-point---both arise from the vanadium $d$-orbitals comprising the kagome lattice.  Given the potential for nesting effects along the M-points at fillings that reach these vHs, slight carrier-doping in these systems is an appealing next step in engineering correlation effects.  

\section{Conclusions} 
The synthesis of single crystals of two new kagome metals GdV$_6$Sn$_6$ and YV$_6$Sn$_6$, each with a nonmagnetic kagome V sublattice, is presented. The lattice structures and electronic ground states were studied via x-ray diffraction, magnetization, magnetotransport and heat capacity measurements. Both compounds possess an ideal $P6/mmm$ symmetry with perfect kagome nets of vanadium atoms coordinated by Sn ions and spaced into layers via interleaving triangular lattice nets of rare earth ions. In the paramgnetic state, DFT modeling categorizes these compounds as $\mathbb{Z}_2$ kagome metals with multiple Dirac crossings and vHs close to E$_F$.  While YV$_6$Sn$_6$ does not show signs of local magnetism, GdV$_6$Sn$_6$ shows signatures of magnetic order below 5 K, and magnetization data collected under low fields suggest the onset of a noncollinear magnetic ground state. Future scattering work will be required to fully determine the zero-field spin structure.  The presence of topological surface states, Dirac points, and vHs's near E$_F$ in the bulk band structure combined with the ability to tune magnetic interactions in these compounds via control of the $R$-sites suggest they are promising platforms for unconventional electronic states born from a model kagome network proximitized with a tunable magnetic layer.

\begin{acknowledgments}
This work was supported via the UC Santa Barbara NSF Quantum Foundry funded via the Q-AMASE-i program under award DMR-1906325. We acknowledge the use of the computing facilities of the Center for Scientific Computing at UC Santa Barbara supported by NSF CNS 1725797 and NSF DMR 1720256. SMLT has been supported by the National Science Foundation Graduate Research Fellowship Program under Grant No. DGE-1650114. BRO and PMS acknowledge financial support from the UC, Santa Barbara through the Elings Fellowship. The work at University of Science and Technology of China (USTC) was supported by the USTC start-up fund and Fundamental Research
Funds for the Central Universities (WK3510000008, WK3510000012).  Any opinions, findings, and conclusions or recommendations expressed in this material are those of the authors and do not necessarily reflect the views of the National Science Foundation. 
\end{acknowledgments}

\bibliography{RV6Sn6.bib}

\begin{thebibliography}{54}%
\makeatletter
\providecommand \@ifxundefined [1]{%
 \@ifx{#1\undefined}
}%
\providecommand \@ifnum [1]{%
 \ifnum #1\expandafter \@firstoftwo
 \else \expandafter \@secondoftwo
 \fi
}%
\providecommand \@ifx [1]{%
 \ifx #1\expandafter \@firstoftwo
 \else \expandafter \@secondoftwo
 \fi
}%
\providecommand \natexlab [1]{#1}%
\providecommand \enquote  [1]{``#1''}%
\providecommand \bibnamefont  [1]{#1}%
\providecommand \bibfnamefont [1]{#1}%
\providecommand \citenamefont [1]{#1}%
\providecommand \href@noop [0]{\@secondoftwo}%
\providecommand \href [0]{\begingroup \@sanitize@url \@href}%
\providecommand \@href[1]{\@@startlink{#1}\@@href}%
\providecommand \@@href[1]{\endgroup#1\@@endlink}%
\providecommand \@sanitize@url [0]{\catcode `\\12\catcode `\$12\catcode
  `\&12\catcode `\#12\catcode `\^12\catcode `\_12\catcode `\%12\relax}%
\providecommand \@@startlink[1]{}%
\providecommand \@@endlink[0]{}%
\providecommand \url  [0]{\begingroup\@sanitize@url \@url }%
\providecommand \@url [1]{\endgroup\@href {#1}{\urlprefix }}%
\providecommand \urlprefix  [0]{URL }%
\providecommand \Eprint [0]{\href }%
\providecommand \doibase [0]{https://doi.org/}%
\providecommand \selectlanguage [0]{\@gobble}%
\providecommand \bibinfo  [0]{\@secondoftwo}%
\providecommand \bibfield  [0]{\@secondoftwo}%
\providecommand \translation [1]{[#1]}%
\providecommand \BibitemOpen [0]{}%
\providecommand \bibitemStop [0]{}%
\providecommand \bibitemNoStop [0]{.\EOS\space}%
\providecommand \EOS [0]{\spacefactor3000\relax}%
\providecommand \BibitemShut  [1]{\csname bibitem#1\endcsname}%
\let\auto@bib@innerbib\@empty
\bibitem [{\citenamefont {Kiesel}\ \emph {et~al.}(2013)\citenamefont {Kiesel},
  \citenamefont {Platt},\ and\ \citenamefont
  {Thomale}}]{PhysRevLett.110.126405}%
  \BibitemOpen
  \bibfield  {author} {\bibinfo {author} {\bibfnamefont {M.~L.}\ \bibnamefont
  {Kiesel}}, \bibinfo {author} {\bibfnamefont {C.}~\bibnamefont {Platt}},\ and\
  \bibinfo {author} {\bibfnamefont {R.}~\bibnamefont {Thomale}},\ }\bibfield
  {title} {\bibinfo {title} {Unconventional fermi surface instabilities in the
  kagome hubbard model},\ }\href
  {https://doi.org/10.1103/PhysRevLett.110.126405} {\bibfield  {journal}
  {\bibinfo  {journal} {Phys. Rev. Lett.}\ }\textbf {\bibinfo {volume} {110}},\
  \bibinfo {pages} {126405} (\bibinfo {year} {2013})}\BibitemShut {NoStop}%
\bibitem [{\citenamefont {Wang}\ \emph {et~al.}(2013)\citenamefont {Wang},
  \citenamefont {Li}, \citenamefont {Xiang},\ and\ \citenamefont
  {Wang}}]{PhysRevB.87.115135}%
  \BibitemOpen
  \bibfield  {author} {\bibinfo {author} {\bibfnamefont {W.-S.}\ \bibnamefont
  {Wang}}, \bibinfo {author} {\bibfnamefont {Z.-Z.}\ \bibnamefont {Li}},
  \bibinfo {author} {\bibfnamefont {Y.-Y.}\ \bibnamefont {Xiang}},\ and\
  \bibinfo {author} {\bibfnamefont {Q.-H.}\ \bibnamefont {Wang}},\ }\bibfield
  {title} {\bibinfo {title} {Competing electronic orders on kagome lattices at
  van hove filling},\ }\href {https://doi.org/10.1103/PhysRevB.87.115135}
  {\bibfield  {journal} {\bibinfo  {journal} {Phys. Rev. B}\ }\textbf {\bibinfo
  {volume} {87}},\ \bibinfo {pages} {115135} (\bibinfo {year}
  {2013})}\BibitemShut {NoStop}%
\bibitem [{\citenamefont {Isakov}\ \emph {et~al.}(2006)\citenamefont {Isakov},
  \citenamefont {Wessel}, \citenamefont {Melko}, \citenamefont {Sengupta},\
  and\ \citenamefont {Kim}}]{PhysRevLett.97.147202}%
  \BibitemOpen
  \bibfield  {author} {\bibinfo {author} {\bibfnamefont {S.~V.}\ \bibnamefont
  {Isakov}}, \bibinfo {author} {\bibfnamefont {S.}~\bibnamefont {Wessel}},
  \bibinfo {author} {\bibfnamefont {R.~G.}\ \bibnamefont {Melko}}, \bibinfo
  {author} {\bibfnamefont {K.}~\bibnamefont {Sengupta}},\ and\ \bibinfo
  {author} {\bibfnamefont {Y.~B.}\ \bibnamefont {Kim}},\ }\bibfield  {title}
  {\bibinfo {title} {Hard-core bosons on the kagome lattice: Valence-bond
  solids and their quantum melting},\ }\href
  {https://doi.org/10.1103/PhysRevLett.97.147202} {\bibfield  {journal}
  {\bibinfo  {journal} {Phys. Rev. Lett.}\ }\textbf {\bibinfo {volume} {97}},\
  \bibinfo {pages} {147202} (\bibinfo {year} {2006})}\BibitemShut {NoStop}%
\bibitem [{\citenamefont {Ko}\ \emph {et~al.}(2009)\citenamefont {Ko},
  \citenamefont {Lee},\ and\ \citenamefont {Wen}}]{ko2009doped}%
  \BibitemOpen
  \bibfield  {author} {\bibinfo {author} {\bibfnamefont {W.-H.}\ \bibnamefont
  {Ko}}, \bibinfo {author} {\bibfnamefont {P.~A.}\ \bibnamefont {Lee}},\ and\
  \bibinfo {author} {\bibfnamefont {X.-G.}\ \bibnamefont {Wen}},\ }\bibfield
  {title} {\bibinfo {title} {Doped kagome system as exotic superconductor},\
  }\href@noop {} {\bibfield  {journal} {\bibinfo  {journal} {Phys. Rev. B}\
  }\textbf {\bibinfo {volume} {79}},\ \bibinfo {pages} {214502} (\bibinfo
  {year} {2009})}\BibitemShut {NoStop}%
\bibitem [{\citenamefont {Kiesel}\ and\ \citenamefont
  {Thomale}(2012)}]{PhysRevB.86.121105}%
  \BibitemOpen
  \bibfield  {author} {\bibinfo {author} {\bibfnamefont {M.~L.}\ \bibnamefont
  {Kiesel}}\ and\ \bibinfo {author} {\bibfnamefont {R.}~\bibnamefont
  {Thomale}},\ }\bibfield  {title} {\bibinfo {title} {Sublattice interference
  in the kagome hubbard model},\ }\href
  {https://doi.org/10.1103/PhysRevB.86.121105} {\bibfield  {journal} {\bibinfo
  {journal} {Phys. Rev. B}\ }\textbf {\bibinfo {volume} {86}},\ \bibinfo
  {pages} {121105} (\bibinfo {year} {2012})}\BibitemShut {NoStop}%
\bibitem [{\citenamefont {Guo}\ and\ \citenamefont
  {Franz}(2009)}]{PhysRevB.80.113102}%
  \BibitemOpen
  \bibfield  {author} {\bibinfo {author} {\bibfnamefont {H.-M.}\ \bibnamefont
  {Guo}}\ and\ \bibinfo {author} {\bibfnamefont {M.}~\bibnamefont {Franz}},\
  }\bibfield  {title} {\bibinfo {title} {Topological insulator on the kagome
  lattice},\ }\href {https://doi.org/10.1103/PhysRevB.80.113102} {\bibfield
  {journal} {\bibinfo  {journal} {Phys. Rev. B}\ }\textbf {\bibinfo {volume}
  {80}},\ \bibinfo {pages} {113102} (\bibinfo {year} {2009})}\BibitemShut
  {NoStop}%
\bibitem [{\citenamefont {Lin}\ and\ \citenamefont
  {Nandkishore}(2021)}]{PhysRevB.104.045122}%
  \BibitemOpen
  \bibfield  {author} {\bibinfo {author} {\bibfnamefont {Y.-P.}\ \bibnamefont
  {Lin}}\ and\ \bibinfo {author} {\bibfnamefont {R.~M.}\ \bibnamefont
  {Nandkishore}},\ }\bibfield  {title} {\bibinfo {title} {Complex charge
  density waves at van hove singularity on hexagonal lattices: Haldane-model
  phase diagram and potential realization in the kagome metals
  $a{V}_{3}{\mathrm{sb}}_{5}$ ($a$=k, rb, cs)},\ }\href
  {https://doi.org/10.1103/PhysRevB.104.045122} {\bibfield  {journal} {\bibinfo
   {journal} {Phys. Rev. B}\ }\textbf {\bibinfo {volume} {104}},\ \bibinfo
  {pages} {045122} (\bibinfo {year} {2021})}\BibitemShut {NoStop}%
\bibitem [{\citenamefont {Yang}\ \emph {et~al.}(2020)\citenamefont {Yang},
  \citenamefont {Wang}, \citenamefont {Ortiz}, \citenamefont {Liu},
  \citenamefont {Gayles}, \citenamefont {Derunova}, \citenamefont
  {Gonzalez-Hernandez}, \citenamefont {{\v S}mejkal}, \citenamefont {Chen},
  \citenamefont {Parkin}, \citenamefont {Wilson}, \citenamefont {Toberer},
  \citenamefont {McQueen},\ and\ \citenamefont {Ali}}]{Yangeabb6003}%
  \BibitemOpen
  \bibfield  {author} {\bibinfo {author} {\bibfnamefont {S.-Y.}\ \bibnamefont
  {Yang}}, \bibinfo {author} {\bibfnamefont {Y.}~\bibnamefont {Wang}}, \bibinfo
  {author} {\bibfnamefont {B.~R.}\ \bibnamefont {Ortiz}}, \bibinfo {author}
  {\bibfnamefont {D.}~\bibnamefont {Liu}}, \bibinfo {author} {\bibfnamefont
  {J.}~\bibnamefont {Gayles}}, \bibinfo {author} {\bibfnamefont
  {E.}~\bibnamefont {Derunova}}, \bibinfo {author} {\bibfnamefont
  {R.}~\bibnamefont {Gonzalez-Hernandez}}, \bibinfo {author} {\bibfnamefont
  {L.}~\bibnamefont {{\v S}mejkal}}, \bibinfo {author} {\bibfnamefont
  {Y.}~\bibnamefont {Chen}}, \bibinfo {author} {\bibfnamefont {S.~S.~P.}\
  \bibnamefont {Parkin}}, \bibinfo {author} {\bibfnamefont {S.~D.}\
  \bibnamefont {Wilson}}, \bibinfo {author} {\bibfnamefont {E.~S.}\
  \bibnamefont {Toberer}}, \bibinfo {author} {\bibfnamefont {T.}~\bibnamefont
  {McQueen}},\ and\ \bibinfo {author} {\bibfnamefont {M.~N.}\ \bibnamefont
  {Ali}},\ }\bibfield  {title} {\bibinfo {title} {Giant, unconventional
  anomalous hall effect in the metallic frustrated magnet candidate,
  \text{KV$_3$Sb$_5$}},\ }\href {https://doi.org/10.1126/sciadv.abb6003}
  {\bibfield  {journal} {\bibinfo  {journal} {Sci. Adv.}\ }\textbf {\bibinfo
  {volume} {6}},\ \bibinfo {pages} {31} (\bibinfo {year} {2020})}\BibitemShut
  {NoStop}%
\bibitem [{\citenamefont {Kida}\ \emph {et~al.}(2011)\citenamefont {Kida},
  \citenamefont {Fenner}, \citenamefont {Dee}, \citenamefont {Terasaki},
  \citenamefont {Hagiwara},\ and\ \citenamefont {Wills}}]{Kida_2011}%
  \BibitemOpen
  \bibfield  {author} {\bibinfo {author} {\bibfnamefont {T.}~\bibnamefont
  {Kida}}, \bibinfo {author} {\bibfnamefont {L.~A.}\ \bibnamefont {Fenner}},
  \bibinfo {author} {\bibfnamefont {A.~A.}\ \bibnamefont {Dee}}, \bibinfo
  {author} {\bibfnamefont {I.}~\bibnamefont {Terasaki}}, \bibinfo {author}
  {\bibfnamefont {M.}~\bibnamefont {Hagiwara}},\ and\ \bibinfo {author}
  {\bibfnamefont {A.~S.}\ \bibnamefont {Wills}},\ }\bibfield  {title} {\bibinfo
  {title} {The giant anomalous hall effect in the ferromagnet
  \text{Fe$_3$Sn$_2$} a frustrated kagome metal},\ }\href
  {https://doi.org/10.1088/0953-8984/23/11/112205} {\bibfield  {journal}
  {\bibinfo  {journal} {J. Phys.: Condens. Matter}\ }\textbf {\bibinfo {volume}
  {23}},\ \bibinfo {pages} {112205} (\bibinfo {year} {2011})}\BibitemShut
  {NoStop}%
\bibitem [{\citenamefont {Liu}\ \emph {et~al.}(2018)\citenamefont {Liu},
  \citenamefont {Sun}, \citenamefont {Kumar}, \citenamefont {Muechler},
  \citenamefont {Sun}, \citenamefont {Jiao}, \citenamefont {Yang},
  \citenamefont {Liu}, \citenamefont {Liang}, \citenamefont {Xu},\ and\
  \citenamefont {et~al.}}]{liu_sun_2018}%
  \BibitemOpen
  \bibfield  {author} {\bibinfo {author} {\bibfnamefont {E.}~\bibnamefont
  {Liu}}, \bibinfo {author} {\bibfnamefont {Y.}~\bibnamefont {Sun}}, \bibinfo
  {author} {\bibfnamefont {N.}~\bibnamefont {Kumar}}, \bibinfo {author}
  {\bibfnamefont {L.}~\bibnamefont {Muechler}}, \bibinfo {author}
  {\bibfnamefont {A.}~\bibnamefont {Sun}}, \bibinfo {author} {\bibfnamefont
  {L.}~\bibnamefont {Jiao}}, \bibinfo {author} {\bibfnamefont {S.-Y.}\
  \bibnamefont {Yang}}, \bibinfo {author} {\bibfnamefont {D.}~\bibnamefont
  {Liu}}, \bibinfo {author} {\bibfnamefont {A.}~\bibnamefont {Liang}}, \bibinfo
  {author} {\bibfnamefont {Q.}~\bibnamefont {Xu}},\ and\ \bibinfo {author}
  {\bibnamefont {et~al.}},\ }\bibfield  {title} {\bibinfo {title} {Giant
  anomalous hall effect in a ferromagnetic kagome-lattice semimetal},\ }\href
  {https://doi.org/10.1038/s41567-018-0234-5} {\bibfield  {journal} {\bibinfo
  {journal} {Nat. Phys.}\ }\textbf {\bibinfo {volume} {14}},\ \bibinfo {pages}
  {1125–1131} (\bibinfo {year} {2018})}\BibitemShut {NoStop}%
\bibitem [{\citenamefont {Fenner}\ \emph {et~al.}(2009)\citenamefont {Fenner},
  \citenamefont {Dee},\ and\ \citenamefont {Wills}}]{Fenner_2009}%
  \BibitemOpen
  \bibfield  {author} {\bibinfo {author} {\bibfnamefont {L.~A.}\ \bibnamefont
  {Fenner}}, \bibinfo {author} {\bibfnamefont {A.~A.}\ \bibnamefont {Dee}},\
  and\ \bibinfo {author} {\bibfnamefont {A.~S.}\ \bibnamefont {Wills}},\
  }\bibfield  {title} {\bibinfo {title} {Non-collinearity and spin frustration
  in the itinerant kagome ferromagnet \text{Fe$_3$Sn$_2$}},\ }\href
  {https://doi.org/10.1088/0953-8984/21/45/452202} {\bibfield  {journal}
  {\bibinfo  {journal} {J. Phys.: Condens. Matter}\ }\textbf {\bibinfo {volume}
  {21}},\ \bibinfo {pages} {452202} (\bibinfo {year} {2009})}\BibitemShut
  {NoStop}%
\bibitem [{\citenamefont {Dally}\ \emph {et~al.}(2021)\citenamefont {Dally},
  \citenamefont {Lynn}, \citenamefont {Ghimire}, \citenamefont {Michel},
  \citenamefont {Siegfried},\ and\ \citenamefont {Mazin}}]{Rebacca_2021}%
  \BibitemOpen
  \bibfield  {author} {\bibinfo {author} {\bibfnamefont {R.~L.}\ \bibnamefont
  {Dally}}, \bibinfo {author} {\bibfnamefont {J.~W.}\ \bibnamefont {Lynn}},
  \bibinfo {author} {\bibfnamefont {N.~J.}\ \bibnamefont {Ghimire}}, \bibinfo
  {author} {\bibfnamefont {D.}~\bibnamefont {Michel}}, \bibinfo {author}
  {\bibfnamefont {P.}~\bibnamefont {Siegfried}},\ and\ \bibinfo {author}
  {\bibfnamefont {I.~I.}\ \bibnamefont {Mazin}},\ }\bibfield  {title} {\bibinfo
  {title} {Chiral properties of the zero-field spiral state and field-induced
  magnetic phases of the itinerant kagome metal
  \text{${\mathrm{YMn}}_{6}{\mathrm{Sn}}_{6}$}},\ }\href
  {https://doi.org/10.1103/PhysRevB.103.094413} {\bibfield  {journal} {\bibinfo
   {journal} {Phys. Rev. B}\ }\textbf {\bibinfo {volume} {103}},\ \bibinfo
  {pages} {094413} (\bibinfo {year} {2021})}\BibitemShut {NoStop}%
\bibitem [{\citenamefont {Jiang}\ \emph {et~al.}(2021)\citenamefont {Jiang},
  \citenamefont {Yin}, \citenamefont {Denner}, \citenamefont {Shumiya},
  \citenamefont {Ortiz}, \citenamefont {Xu}, \citenamefont {Guguchia},
  \citenamefont {He}, \citenamefont {Hossain}, \citenamefont {Liu},
  \citenamefont {Ruff}, \citenamefont {Kautzsch}, \citenamefont {Zhang},
  \citenamefont {Chang}, \citenamefont {Belopolski}, \citenamefont {Zhang},
  \citenamefont {Cochran}, \citenamefont {Multer}, \citenamefont {Litskevich},
  \citenamefont {Cheng}, \citenamefont {Yang}, \citenamefont {Wang},
  \citenamefont {Thomale}, \citenamefont {Neupert}, \citenamefont {Wilson},\
  and\ \citenamefont {Hasan}}]{Jiang2021}%
  \BibitemOpen
  \bibfield  {author} {\bibinfo {author} {\bibfnamefont {Y.-X.}\ \bibnamefont
  {Jiang}}, \bibinfo {author} {\bibfnamefont {J.-X.}\ \bibnamefont {Yin}},
  \bibinfo {author} {\bibfnamefont {M.~M.}\ \bibnamefont {Denner}}, \bibinfo
  {author} {\bibfnamefont {N.}~\bibnamefont {Shumiya}}, \bibinfo {author}
  {\bibfnamefont {B.~R.}\ \bibnamefont {Ortiz}}, \bibinfo {author}
  {\bibfnamefont {G.}~\bibnamefont {Xu}}, \bibinfo {author} {\bibfnamefont
  {Z.}~\bibnamefont {Guguchia}}, \bibinfo {author} {\bibfnamefont
  {J.}~\bibnamefont {He}}, \bibinfo {author} {\bibfnamefont {M.~S.}\
  \bibnamefont {Hossain}}, \bibinfo {author} {\bibfnamefont {X.}~\bibnamefont
  {Liu}}, \bibinfo {author} {\bibfnamefont {J.}~\bibnamefont {Ruff}}, \bibinfo
  {author} {\bibfnamefont {L.}~\bibnamefont {Kautzsch}}, \bibinfo {author}
  {\bibfnamefont {S.~S.}\ \bibnamefont {Zhang}}, \bibinfo {author}
  {\bibfnamefont {G.}~\bibnamefont {Chang}}, \bibinfo {author} {\bibfnamefont
  {I.}~\bibnamefont {Belopolski}}, \bibinfo {author} {\bibfnamefont
  {Q.}~\bibnamefont {Zhang}}, \bibinfo {author} {\bibfnamefont {T.~A.}\
  \bibnamefont {Cochran}}, \bibinfo {author} {\bibfnamefont {D.}~\bibnamefont
  {Multer}}, \bibinfo {author} {\bibfnamefont {M.}~\bibnamefont {Litskevich}},
  \bibinfo {author} {\bibfnamefont {Z.-J.}\ \bibnamefont {Cheng}}, \bibinfo
  {author} {\bibfnamefont {X.~P.}\ \bibnamefont {Yang}}, \bibinfo {author}
  {\bibfnamefont {Z.}~\bibnamefont {Wang}}, \bibinfo {author} {\bibfnamefont
  {R.}~\bibnamefont {Thomale}}, \bibinfo {author} {\bibfnamefont
  {T.}~\bibnamefont {Neupert}}, \bibinfo {author} {\bibfnamefont {S.~D.}\
  \bibnamefont {Wilson}},\ and\ \bibinfo {author} {\bibfnamefont {M.~Z.}\
  \bibnamefont {Hasan}},\ }\bibfield  {title} {\bibinfo {title} {Unconventional
  chiral charge order in kagome superconductor kv3sb5},\ }\href
  {https://doi.org/10.1038/s41563-021-01034-y} {\bibfield  {journal} {\bibinfo
  {journal} {Nature Materials}\ }\textbf {\bibinfo {volume} {20}},\ \bibinfo
  {pages} {1353} (\bibinfo {year} {2021})}\BibitemShut {NoStop}%
\bibitem [{\citenamefont {Zhao}\ \emph {et~al.}(2021)\citenamefont {Zhao},
  \citenamefont {Li}, \citenamefont {Ortiz}, \citenamefont {Teicher},
  \citenamefont {Park}, \citenamefont {Ye}, \citenamefont {Wang}, \citenamefont
  {Balents}, \citenamefont {Wilson},\ and\ \citenamefont
  {Zeljkovic}}]{Zhao2021}%
  \BibitemOpen
  \bibfield  {author} {\bibinfo {author} {\bibfnamefont {H.}~\bibnamefont
  {Zhao}}, \bibinfo {author} {\bibfnamefont {H.}~\bibnamefont {Li}}, \bibinfo
  {author} {\bibfnamefont {B.~R.}\ \bibnamefont {Ortiz}}, \bibinfo {author}
  {\bibfnamefont {S.~M.~L.}\ \bibnamefont {Teicher}}, \bibinfo {author}
  {\bibfnamefont {T.}~\bibnamefont {Park}}, \bibinfo {author} {\bibfnamefont
  {M.}~\bibnamefont {Ye}}, \bibinfo {author} {\bibfnamefont {Z.}~\bibnamefont
  {Wang}}, \bibinfo {author} {\bibfnamefont {L.}~\bibnamefont {Balents}},
  \bibinfo {author} {\bibfnamefont {S.~D.}\ \bibnamefont {Wilson}},\ and\
  \bibinfo {author} {\bibfnamefont {I.}~\bibnamefont {Zeljkovic}},\ }\bibfield
  {title} {\bibinfo {title} {Cascade of correlated electron states in a kagome
  superconductor csv3sb5},\ }\bibfield  {journal} {\bibinfo  {journal}
  {Nature}\ }\href {https://doi.org/10.1038/s41586-021-03946-w}
  {10.1038/s41586-021-03946-w} (\bibinfo {year} {2021})\BibitemShut {NoStop}%
\bibitem [{\citenamefont {Ortiz}\ \emph
  {et~al.}(2020{\natexlab{a}})\citenamefont {Ortiz}, \citenamefont {Kenney},
  \citenamefont {Sarte}, \citenamefont {Teicher}, \citenamefont {Seshadri},
  \citenamefont {Graf},\ and\ \citenamefont
  {Wilson}}]{ortiz2020superconductivity}%
  \BibitemOpen
  \bibfield  {author} {\bibinfo {author} {\bibfnamefont {B.~R.}\ \bibnamefont
  {Ortiz}}, \bibinfo {author} {\bibfnamefont {E.}~\bibnamefont {Kenney}},
  \bibinfo {author} {\bibfnamefont {P.~M.}\ \bibnamefont {Sarte}}, \bibinfo
  {author} {\bibfnamefont {S.~M.}\ \bibnamefont {Teicher}}, \bibinfo {author}
  {\bibfnamefont {R.}~\bibnamefont {Seshadri}}, \bibinfo {author}
  {\bibfnamefont {M.~J.}\ \bibnamefont {Graf}},\ and\ \bibinfo {author}
  {\bibfnamefont {S.~D.}\ \bibnamefont {Wilson}},\ }\bibfield  {title}
  {\bibinfo {title} {{Superconductivity in the \text{$\mathbb{Z}_2$} kagome
  metal \text{KV$_3$Sb$_5$}}},\ }\href@noop {} {\bibfield  {journal} {\bibinfo
  {journal} {Phys. Rev. Mater.}\ }\textbf {\bibinfo {volume} {5}},\ \bibinfo
  {pages} {034801} (\bibinfo {year} {2020}{\natexlab{a}})}\BibitemShut
  {NoStop}%
\bibitem [{\citenamefont {Ortiz}\ \emph
  {et~al.}(2020{\natexlab{b}})\citenamefont {Ortiz}, \citenamefont {Teicher},
  \citenamefont {Hu}, \citenamefont {Zuo}, \citenamefont {Sarte}, \citenamefont
  {Schueller}, \citenamefont {Abeykoon}, \citenamefont {Krogstad},
  \citenamefont {Rosenkranz}, \citenamefont {Osborn}, \citenamefont {Seshadri},
  \citenamefont {Balents}, \citenamefont {He},\ and\ \citenamefont
  {Wilson}}]{ortizCsV3Sb5}%
  \BibitemOpen
  \bibfield  {author} {\bibinfo {author} {\bibfnamefont {B.~R.}\ \bibnamefont
  {Ortiz}}, \bibinfo {author} {\bibfnamefont {S.~M.}\ \bibnamefont {Teicher}},
  \bibinfo {author} {\bibfnamefont {Y.}~\bibnamefont {Hu}}, \bibinfo {author}
  {\bibfnamefont {J.~L.}\ \bibnamefont {Zuo}}, \bibinfo {author} {\bibfnamefont
  {P.~M.}\ \bibnamefont {Sarte}}, \bibinfo {author} {\bibfnamefont {E.~C.}\
  \bibnamefont {Schueller}}, \bibinfo {author} {\bibfnamefont {A.~M.}\
  \bibnamefont {Abeykoon}}, \bibinfo {author} {\bibfnamefont {M.~J.}\
  \bibnamefont {Krogstad}}, \bibinfo {author} {\bibfnamefont {S.}~\bibnamefont
  {Rosenkranz}}, \bibinfo {author} {\bibfnamefont {R.}~\bibnamefont {Osborn}},
  \bibinfo {author} {\bibfnamefont {R.}~\bibnamefont {Seshadri}}, \bibinfo
  {author} {\bibfnamefont {L.}~\bibnamefont {Balents}}, \bibinfo {author}
  {\bibfnamefont {J.}~\bibnamefont {He}},\ and\ \bibinfo {author}
  {\bibfnamefont {S.~D.}\ \bibnamefont {Wilson}},\ }\bibfield  {title}
  {\bibinfo {title} {{CsV$_3$Sb$_5$: a $\mathbb{Z}_2$ topological kagome metal
  with a superconducting ground state}},\ }\href@noop {} {\bibfield  {journal}
  {\bibinfo  {journal} {Phys. Rev. Lett.}\ }\textbf {\bibinfo {volume} {125}},\
  \bibinfo {pages} {247002} (\bibinfo {year} {2020}{\natexlab{b}})}\BibitemShut
  {NoStop}%
\bibitem [{\citenamefont {Yin}\ \emph {et~al.}(2021)\citenamefont {Yin},
  \citenamefont {Tu}, \citenamefont {Gong}, \citenamefont {Fu}, \citenamefont
  {Yan},\ and\ \citenamefont {Lei}}]{2021Rb}%
  \BibitemOpen
  \bibfield  {author} {\bibinfo {author} {\bibfnamefont {Q.}~\bibnamefont
  {Yin}}, \bibinfo {author} {\bibfnamefont {Z.}~\bibnamefont {Tu}}, \bibinfo
  {author} {\bibfnamefont {C.}~\bibnamefont {Gong}}, \bibinfo {author}
  {\bibfnamefont {Y.}~\bibnamefont {Fu}}, \bibinfo {author} {\bibfnamefont
  {S.}~\bibnamefont {Yan}},\ and\ \bibinfo {author} {\bibfnamefont
  {H.}~\bibnamefont {Lei}},\ }\bibfield  {title} {\bibinfo {title}
  {Superconductivity and normal-state properties of kagome metal {RbV}3sb5
  single crystals},\ }\href {https://doi.org/10.1088/0256-307x/38/3/037403} {\
  \textbf {\bibinfo {volume} {38}},\ \bibinfo {pages} {037403} (\bibinfo {year}
  {2021})}\BibitemShut {NoStop}%
\bibitem [{\citenamefont {Li}\ \emph {et~al.}(2021)\citenamefont {Li},
  \citenamefont {Wang}, \citenamefont {Wang}, \citenamefont {Yuan},
  \citenamefont {Song}, \citenamefont {Lou}, \citenamefont {Liu}, \citenamefont
  {Huang}, \citenamefont {Liu}, \citenamefont {Lei}, \citenamefont {Yin},\ and\
  \citenamefont {Wang}}]{Li_2021_dirac}%
  \BibitemOpen
  \bibfield  {author} {\bibinfo {author} {\bibfnamefont {M.}~\bibnamefont
  {Li}}, \bibinfo {author} {\bibfnamefont {Q.}~\bibnamefont {Wang}}, \bibinfo
  {author} {\bibfnamefont {G.}~\bibnamefont {Wang}}, \bibinfo {author}
  {\bibfnamefont {Z.}~\bibnamefont {Yuan}}, \bibinfo {author} {\bibfnamefont
  {W.}~\bibnamefont {Song}}, \bibinfo {author} {\bibfnamefont {R.}~\bibnamefont
  {Lou}}, \bibinfo {author} {\bibfnamefont {Z.}~\bibnamefont {Liu}}, \bibinfo
  {author} {\bibfnamefont {Y.}~\bibnamefont {Huang}}, \bibinfo {author}
  {\bibfnamefont {Z.}~\bibnamefont {Liu}}, \bibinfo {author} {\bibfnamefont
  {H.}~\bibnamefont {Lei}}, \bibinfo {author} {\bibfnamefont {Z.}~\bibnamefont
  {Yin}},\ and\ \bibinfo {author} {\bibfnamefont {S.}~\bibnamefont {Wang}},\
  }\bibfield  {title} {\bibinfo {title} {Spin-polarized dirac cone, flat band
  and saddle point in kagome magnet \text{YMn$_6$Sn$_6$}},\ }\href
  {https://doi.org/10.1038/s41467-021-23536-8} {\bibfield  {journal} {\bibinfo
  {journal} {Nat. Commun.}\ }\textbf {\bibinfo {volume} {12}},\ \bibinfo
  {pages} {3129} (\bibinfo {year} {2021})}\BibitemShut {NoStop}%
\bibitem [{\citenamefont {Chen}\ \emph {et~al.}(2021)\citenamefont {Chen},
  \citenamefont {Le}, \citenamefont {Fu}, \citenamefont {Lin}, \citenamefont
  {Schnelle}, \citenamefont {Sun},\ and\ \citenamefont
  {Felser}}]{Chen_Dong_2021}%
  \BibitemOpen
  \bibfield  {author} {\bibinfo {author} {\bibfnamefont {D.}~\bibnamefont
  {Chen}}, \bibinfo {author} {\bibfnamefont {C.}~\bibnamefont {Le}}, \bibinfo
  {author} {\bibfnamefont {C.}~\bibnamefont {Fu}}, \bibinfo {author}
  {\bibfnamefont {H.}~\bibnamefont {Lin}}, \bibinfo {author} {\bibfnamefont
  {W.}~\bibnamefont {Schnelle}}, \bibinfo {author} {\bibfnamefont
  {Y.}~\bibnamefont {Sun}},\ and\ \bibinfo {author} {\bibfnamefont
  {C.}~\bibnamefont {Felser}},\ }\bibfield  {title} {\bibinfo {title} {Large
  anomalous hall effect in the kagome ferromagnet
  \text{${\mathrm{LiMn}}_{6}{\mathrm{Sn}}_{6}$}},\ }\href
  {https://doi.org/10.1103/PhysRevB.103.144410} {\bibfield  {journal} {\bibinfo
   {journal} {Phys. Rev. B}\ }\textbf {\bibinfo {volume} {103}},\ \bibinfo
  {pages} {144410} (\bibinfo {year} {2021})}\BibitemShut {NoStop}%
\bibitem [{\citenamefont {Asaba}\ \emph {et~al.}(2020)\citenamefont {Asaba},
  \citenamefont {Thomas}, \citenamefont {Curtis}, \citenamefont {Thompson},
  \citenamefont {Bauer},\ and\ \citenamefont {Ronning}}]{Asaba_2020}%
  \BibitemOpen
  \bibfield  {author} {\bibinfo {author} {\bibfnamefont {T.}~\bibnamefont
  {Asaba}}, \bibinfo {author} {\bibfnamefont {S.~M.}\ \bibnamefont {Thomas}},
  \bibinfo {author} {\bibfnamefont {M.}~\bibnamefont {Curtis}}, \bibinfo
  {author} {\bibfnamefont {J.~D.}\ \bibnamefont {Thompson}}, \bibinfo {author}
  {\bibfnamefont {E.~D.}\ \bibnamefont {Bauer}},\ and\ \bibinfo {author}
  {\bibfnamefont {F.}~\bibnamefont {Ronning}},\ }\bibfield  {title} {\bibinfo
  {title} {Anomalous hall effect in the kagome ferrimagnet
  \text{${\mathrm{GdMn}}_{6}{\mathrm{Sn}}_{6}$}},\ }\href
  {https://doi.org/10.1103/PhysRevB.101.174415} {\bibfield  {journal} {\bibinfo
   {journal} {Phys. Rev. B}\ }\textbf {\bibinfo {volume} {101}},\ \bibinfo
  {pages} {174415} (\bibinfo {year} {2020})}\BibitemShut {NoStop}%
\bibitem [{\citenamefont {Yin}\ \emph {et~al.}(2020)\citenamefont {Yin},
  \citenamefont {Ma}, \citenamefont {Cochran}, \citenamefont {Xu},
  \citenamefont {Zhang}, \citenamefont {Tien}, \citenamefont {Shumiya},
  \citenamefont {Cheng}, \citenamefont {Jiang}, \citenamefont {Lian},
  \citenamefont {Song}, \citenamefont {Chang}, \citenamefont {Belopolski},
  \citenamefont {Multer}, \citenamefont {Litskevich}, \citenamefont {Cheng},
  \citenamefont {Yang}, \citenamefont {Swidler}, \citenamefont {Zhou},
  \citenamefont {Lin}, \citenamefont {Neupert}, \citenamefont {Wang},
  \citenamefont {Yao}, \citenamefont {Chang}, \citenamefont {Jia},\ and\
  \citenamefont {Zahid~Hasan}}]{Yin2020}%
  \BibitemOpen
  \bibfield  {author} {\bibinfo {author} {\bibfnamefont {J.-X.}\ \bibnamefont
  {Yin}}, \bibinfo {author} {\bibfnamefont {W.}~\bibnamefont {Ma}}, \bibinfo
  {author} {\bibfnamefont {T.~A.}\ \bibnamefont {Cochran}}, \bibinfo {author}
  {\bibfnamefont {X.}~\bibnamefont {Xu}}, \bibinfo {author} {\bibfnamefont
  {S.~S.}\ \bibnamefont {Zhang}}, \bibinfo {author} {\bibfnamefont {H.-J.}\
  \bibnamefont {Tien}}, \bibinfo {author} {\bibfnamefont {N.}~\bibnamefont
  {Shumiya}}, \bibinfo {author} {\bibfnamefont {G.}~\bibnamefont {Cheng}},
  \bibinfo {author} {\bibfnamefont {K.}~\bibnamefont {Jiang}}, \bibinfo
  {author} {\bibfnamefont {B.}~\bibnamefont {Lian}}, \bibinfo {author}
  {\bibfnamefont {Z.}~\bibnamefont {Song}}, \bibinfo {author} {\bibfnamefont
  {G.}~\bibnamefont {Chang}}, \bibinfo {author} {\bibfnamefont
  {I.}~\bibnamefont {Belopolski}}, \bibinfo {author} {\bibfnamefont
  {D.}~\bibnamefont {Multer}}, \bibinfo {author} {\bibfnamefont
  {M.}~\bibnamefont {Litskevich}}, \bibinfo {author} {\bibfnamefont {Z.-J.}\
  \bibnamefont {Cheng}}, \bibinfo {author} {\bibfnamefont {X.~P.}\ \bibnamefont
  {Yang}}, \bibinfo {author} {\bibfnamefont {B.}~\bibnamefont {Swidler}},
  \bibinfo {author} {\bibfnamefont {H.}~\bibnamefont {Zhou}}, \bibinfo {author}
  {\bibfnamefont {H.}~\bibnamefont {Lin}}, \bibinfo {author} {\bibfnamefont
  {T.}~\bibnamefont {Neupert}}, \bibinfo {author} {\bibfnamefont
  {Z.}~\bibnamefont {Wang}}, \bibinfo {author} {\bibfnamefont {N.}~\bibnamefont
  {Yao}}, \bibinfo {author} {\bibfnamefont {T.-R.}\ \bibnamefont {Chang}},
  \bibinfo {author} {\bibfnamefont {S.}~\bibnamefont {Jia}},\ and\ \bibinfo
  {author} {\bibfnamefont {M.}~\bibnamefont {Zahid~Hasan}},\ }\bibfield
  {title} {\bibinfo {title} {Quantum-limit chern topological magnetism in
  \text{TbMn$_6$Sn$_6$}},\ }\href {https://doi.org/10.1038/s41586-020-2482-7}
  {\bibfield  {journal} {\bibinfo  {journal} {Nature}\ }\textbf {\bibinfo
  {volume} {583}},\ \bibinfo {pages} {533} (\bibinfo {year}
  {2020})}\BibitemShut {NoStop}%
\bibitem [{\citenamefont {Ghimire}\ \emph {et~al.}(2020)\citenamefont
  {Ghimire}, \citenamefont {Dally}, \citenamefont {Poudel}, \citenamefont
  {Jones}, \citenamefont {Michel}, \citenamefont {Magar}, \citenamefont
  {Bleuel}, \citenamefont {McGuire}, \citenamefont {Jiang}, \citenamefont
  {Mitchell}, \citenamefont {Lynn},\ and\ \citenamefont
  {Mazin}}]{Ghimire_2020}%
  \BibitemOpen
  \bibfield  {author} {\bibinfo {author} {\bibfnamefont {N.~J.}\ \bibnamefont
  {Ghimire}}, \bibinfo {author} {\bibfnamefont {R.~L.}\ \bibnamefont {Dally}},
  \bibinfo {author} {\bibfnamefont {L.}~\bibnamefont {Poudel}}, \bibinfo
  {author} {\bibfnamefont {D.~C.}\ \bibnamefont {Jones}}, \bibinfo {author}
  {\bibfnamefont {D.}~\bibnamefont {Michel}}, \bibinfo {author} {\bibfnamefont
  {N.~T.}\ \bibnamefont {Magar}}, \bibinfo {author} {\bibfnamefont
  {M.}~\bibnamefont {Bleuel}}, \bibinfo {author} {\bibfnamefont {M.~A.}\
  \bibnamefont {McGuire}}, \bibinfo {author} {\bibfnamefont {J.~S.}\
  \bibnamefont {Jiang}}, \bibinfo {author} {\bibfnamefont {J.~F.}\ \bibnamefont
  {Mitchell}}, \bibinfo {author} {\bibfnamefont {J.~W.}\ \bibnamefont {Lynn}},\
  and\ \bibinfo {author} {\bibfnamefont {I.~I.}\ \bibnamefont {Mazin}},\
  }\bibfield  {title} {\bibinfo {title} {Competing magnetic phases and
  fluctuation-driven scalar spin chirality in the kagome metal
  \text{YMn$_6$Sn$_6$}},\ }\href {https://doi.org/10.1126/sciadv.abe2680}
  {\bibfield  {journal} {\bibinfo  {journal} {Sci. Adv.}\ }\textbf {\bibinfo
  {volume} {6}},\ \bibinfo {pages} {51} (\bibinfo {year} {2020})}\BibitemShut
  {NoStop}%
\bibitem [{\citenamefont {Gieck}\ \emph {et~al.}(2006)\citenamefont {Gieck},
  \citenamefont {Schreyer}, \citenamefont {Fässler}, \citenamefont {Cavet},\
  and\ \citenamefont {Claus}}]{Gieck_2006}%
  \BibitemOpen
  \bibfield  {author} {\bibinfo {author} {\bibfnamefont {C.}~\bibnamefont
  {Gieck}}, \bibinfo {author} {\bibfnamefont {M.}~\bibnamefont {Schreyer}},
  \bibinfo {author} {\bibfnamefont {T.~F.}\ \bibnamefont {Fässler}}, \bibinfo
  {author} {\bibfnamefont {S.}~\bibnamefont {Cavet}},\ and\ \bibinfo {author}
  {\bibfnamefont {P.}~\bibnamefont {Claus}},\ }\bibfield  {title} {\bibinfo
  {title} {Synthesis, crystal structure, and catalytic properties of
  \text{MgCo$_6$Ge$_6$}},\ }\href
  {https://doi.org/https://doi.org/10.1002/chem.200500411} {\bibfield
  {journal} {\bibinfo  {journal} {Chem. Eur. J}\ }\textbf {\bibinfo {volume}
  {12}},\ \bibinfo {pages} {1924} (\bibinfo {year} {2006})}\BibitemShut
  {NoStop}%
\bibitem [{\citenamefont {Zhang}\ \emph {et~al.}(2001)\citenamefont {Zhang},
  \citenamefont {Zhao}, \citenamefont {Cheng}, \citenamefont {Li},
  \citenamefont {Sun}, \citenamefont {Zhang},\ and\ \citenamefont
  {Shen}}]{Zhang_2001}%
  \BibitemOpen
  \bibfield  {author} {\bibinfo {author} {\bibfnamefont {S.-y.}\ \bibnamefont
  {Zhang}}, \bibinfo {author} {\bibfnamefont {P.}~\bibnamefont {Zhao}},
  \bibinfo {author} {\bibfnamefont {Z.-h.}\ \bibnamefont {Cheng}}, \bibinfo
  {author} {\bibfnamefont {R.-w.}\ \bibnamefont {Li}}, \bibinfo {author}
  {\bibfnamefont {J.-r.}\ \bibnamefont {Sun}}, \bibinfo {author} {\bibfnamefont
  {H.-w.}\ \bibnamefont {Zhang}},\ and\ \bibinfo {author} {\bibfnamefont
  {B.-g.}\ \bibnamefont {Shen}},\ }\bibfield  {title} {\bibinfo {title}
  {Magnetism and giant magnetoresistance of
  \text{${\mathrm{YMn}}_{6}{\mathrm{Sn}}_{6\ensuremath{-}x}{\mathrm{Ga}}_{x}
  (x=0-1.8)$} compounds},\ }\href {https://doi.org/10.1103/PhysRevB.64.212404}
  {\bibfield  {journal} {\bibinfo  {journal} {Phys. Rev. B}\ }\textbf {\bibinfo
  {volume} {64}},\ \bibinfo {pages} {212404} (\bibinfo {year}
  {2001})}\BibitemShut {NoStop}%
\bibitem [{\citenamefont {Schobinger-Papamantellos}\ \emph
  {et~al.}(2016)\citenamefont {Schobinger-Papamantellos}, \citenamefont
  {Rodr{\'\i}guez-Carvajal},\ and\ \citenamefont {Buschow}}]{SCHOBINGER_2016}%
  \BibitemOpen
  \bibfield  {author} {\bibinfo {author} {\bibfnamefont {P.}~\bibnamefont
  {Schobinger-Papamantellos}}, \bibinfo {author} {\bibfnamefont
  {J.}~\bibnamefont {Rodr{\'\i}guez-Carvajal}},\ and\ \bibinfo {author}
  {\bibfnamefont {K.}~\bibnamefont {Buschow}},\ }\bibfield  {title} {\bibinfo
  {title} {Cycloid spirals and cycloid cone transition in the
  \text{HoMn$_{6-x}$Cr$_x$Ge$_6$ (T, x)} magnetic phase diagramm by neutron
  diffraction},\ }\href@noop {} {\bibfield  {journal} {\bibinfo  {journal} {J.
  Magn. Magn. Mater.}\ }\textbf {\bibinfo {volume} {408}},\ \bibinfo {pages}
  {233} (\bibinfo {year} {2016})}\BibitemShut {NoStop}%
\bibitem [{\citenamefont {Ortiz}\ \emph {et~al.}(2019)\citenamefont {Ortiz},
  \citenamefont {Gomes}, \citenamefont {Morey}, \citenamefont {Winiarski},
  \citenamefont {Bordelon}, \citenamefont {Mangum}, \citenamefont {Oswald},
  \citenamefont {Rodriguez-Rivera}, \citenamefont {Neilson}, \citenamefont
  {Wilson}, \citenamefont {Ertekin}, \citenamefont {McQueen},\ and\
  \citenamefont {Toberer}}]{Brenden_2019}%
  \BibitemOpen
  \bibfield  {author} {\bibinfo {author} {\bibfnamefont {B.~R.}\ \bibnamefont
  {Ortiz}}, \bibinfo {author} {\bibfnamefont {L.~C.}\ \bibnamefont {Gomes}},
  \bibinfo {author} {\bibfnamefont {J.~R.}\ \bibnamefont {Morey}}, \bibinfo
  {author} {\bibfnamefont {M.}~\bibnamefont {Winiarski}}, \bibinfo {author}
  {\bibfnamefont {M.}~\bibnamefont {Bordelon}}, \bibinfo {author}
  {\bibfnamefont {J.~S.}\ \bibnamefont {Mangum}}, \bibinfo {author}
  {\bibfnamefont {I.~W.~H.}\ \bibnamefont {Oswald}}, \bibinfo {author}
  {\bibfnamefont {J.~A.}\ \bibnamefont {Rodriguez-Rivera}}, \bibinfo {author}
  {\bibfnamefont {J.~R.}\ \bibnamefont {Neilson}}, \bibinfo {author}
  {\bibfnamefont {S.~D.}\ \bibnamefont {Wilson}}, \bibinfo {author}
  {\bibfnamefont {E.}~\bibnamefont {Ertekin}}, \bibinfo {author} {\bibfnamefont
  {T.~M.}\ \bibnamefont {McQueen}},\ and\ \bibinfo {author} {\bibfnamefont
  {E.~S.}\ \bibnamefont {Toberer}},\ }\bibfield  {title} {\bibinfo {title} {New
  kagome prototype materials: discovery of
  \text{${\mathrm{KV}}_{3}{\mathrm{Sb}}_{5},{\mathrm{RbV}}_{3}{\mathrm{Sb}}_{5}$},
  and \text{${\mathrm{CsV}}_{3}{\mathrm{Sb}}_{5}$}},\ }\href
  {https://doi.org/10.1103/PhysRevMaterials.3.094407} {\bibfield  {journal}
  {\bibinfo  {journal} {Phys. Rev. Materials}\ }\textbf {\bibinfo {volume}
  {3}},\ \bibinfo {pages} {094407} (\bibinfo {year} {2019})}\BibitemShut
  {NoStop}%
\bibitem [{\citenamefont {Jiang}\ \emph {et~al.}(2020)\citenamefont {Jiang},
  \citenamefont {Yin}, \citenamefont {Denner}, \citenamefont {Shumiya},
  \citenamefont {Ortiz}, \citenamefont {He}, \citenamefont {Liu}, \citenamefont
  {Zhang}, \citenamefont {Chang}, \citenamefont {Belopolski} \emph
  {et~al.}}]{yuxiaoKVS}%
  \BibitemOpen
  \bibfield  {author} {\bibinfo {author} {\bibfnamefont {Y.-X.}\ \bibnamefont
  {Jiang}}, \bibinfo {author} {\bibfnamefont {J.-X.}\ \bibnamefont {Yin}},
  \bibinfo {author} {\bibfnamefont {M.~M.}\ \bibnamefont {Denner}}, \bibinfo
  {author} {\bibfnamefont {N.}~\bibnamefont {Shumiya}}, \bibinfo {author}
  {\bibfnamefont {B.~R.}\ \bibnamefont {Ortiz}}, \bibinfo {author}
  {\bibfnamefont {J.}~\bibnamefont {He}}, \bibinfo {author} {\bibfnamefont
  {X.}~\bibnamefont {Liu}}, \bibinfo {author} {\bibfnamefont {S.~S.}\
  \bibnamefont {Zhang}}, \bibinfo {author} {\bibfnamefont {G.}~\bibnamefont
  {Chang}}, \bibinfo {author} {\bibfnamefont {I.}~\bibnamefont {Belopolski}},
  \emph {et~al.},\ }\bibfield  {title} {\bibinfo {title} {{Discovery of
  topological charge order in kagome superconductor KV$_3$Sb$_5$}},\
  }\href@noop {} {\bibfield  {journal} {\bibinfo  {journal} {arXiv:2012.15709}\
  } (\bibinfo {year} {2020})}\BibitemShut {NoStop}%
\bibitem [{\citenamefont {Sheldrick}(2015)}]{Shelx_2015}%
  \BibitemOpen
  \bibfield  {author} {\bibinfo {author} {\bibfnamefont {G.~M.}\ \bibnamefont
  {Sheldrick}},\ }\bibfield  {title} {\bibinfo {title} {{{\it SHELXT} {--}
  Integrated space-group and crystal-structure determination}},\ }\href
  {https://doi.org/10.1107/S2053273314026370} {\bibfield  {journal} {\bibinfo
  {journal} {Acta Cryst. A}\ }\textbf {\bibinfo {volume} {71}},\ \bibinfo
  {pages} {3} (\bibinfo {year} {2015})}\BibitemShut {NoStop}%
\bibitem [{\citenamefont {Kresse}\ and\ \citenamefont
  {Hafner}(1994)}]{Kresse1994}%
  \BibitemOpen
  \bibfield  {author} {\bibinfo {author} {\bibfnamefont {G.}~\bibnamefont
  {Kresse}}\ and\ \bibinfo {author} {\bibfnamefont {J.}~\bibnamefont
  {Hafner}},\ }\bibfield  {title} {\bibinfo {title} {Ab initio
  molecular-dynamics simulation of the liquid-metal--amorphous-semiconductor
  transition in germanium},\ }\href {https://doi.org/10.1103/PhysRevB.49.14251}
  {\bibfield  {journal} {\bibinfo  {journal} {Phys. Rev. B}\ }\textbf {\bibinfo
  {volume} {49}},\ \bibinfo {pages} {14251} (\bibinfo {year}
  {1994})}\BibitemShut {NoStop}%
\bibitem [{\citenamefont {Kresse}\ and\ \citenamefont
  {Furthm{\"u}ller}(1996{\natexlab{a}})}]{Kresse1996a}%
  \BibitemOpen
  \bibfield  {author} {\bibinfo {author} {\bibfnamefont {G.}~\bibnamefont
  {Kresse}}\ and\ \bibinfo {author} {\bibfnamefont {J.}~\bibnamefont
  {Furthm{\"u}ller}},\ }\bibfield  {title} {\bibinfo {title} {Efficient
  iterative schemes for ab initio total-energy calculations using a plane-wave
  basis set},\ }\href {https://doi.org/10.1103/PhysRevB.54.11169} {\bibfield
  {journal} {\bibinfo  {journal} {Phys. Rev. B}\ }\textbf {\bibinfo {volume}
  {54}},\ \bibinfo {pages} {11169} (\bibinfo {year}
  {1996}{\natexlab{a}})}\BibitemShut {NoStop}%
\bibitem [{\citenamefont {Kresse}\ and\ \citenamefont
  {Furthm{\"u}ller}(1996{\natexlab{b}})}]{Kresse1996b}%
  \BibitemOpen
  \bibfield  {author} {\bibinfo {author} {\bibfnamefont {G.}~\bibnamefont
  {Kresse}}\ and\ \bibinfo {author} {\bibfnamefont {J.}~\bibnamefont
  {Furthm{\"u}ller}},\ }\bibfield  {title} {\bibinfo {title} {Efficiency of
  ab-initio total energy calculations for metals and semiconductors using a
  plane-wave basis set},\ }\href {https://doi.org/10.1016/0927-0256(96)00008-0}
  {\bibfield  {journal} {\bibinfo  {journal} {Comput. Mater. Sci.}\ }\textbf
  {\bibinfo {volume} {6}},\ \bibinfo {pages} {15} (\bibinfo {year}
  {1996}{\natexlab{b}})}\BibitemShut {NoStop}%
\bibitem [{\citenamefont {Perdew}\ \emph {et~al.}(1996)\citenamefont {Perdew},
  \citenamefont {Burke},\ and\ \citenamefont {Ernzerhof}}]{Perdew1996}%
  \BibitemOpen
  \bibfield  {author} {\bibinfo {author} {\bibfnamefont {J.~P.}\ \bibnamefont
  {Perdew}}, \bibinfo {author} {\bibfnamefont {K.}~\bibnamefont {Burke}},\ and\
  \bibinfo {author} {\bibfnamefont {M.}~\bibnamefont {Ernzerhof}},\ }\bibfield
  {title} {\bibinfo {title} {Generalized gradient approximation made simple},\
  }\href {https://doi.org/10.1103/PhysRevLett.77.3865} {\bibfield  {journal}
  {\bibinfo  {journal} {Phys. Rev. Lett.}\ }\textbf {\bibinfo {volume} {77}},\
  \bibinfo {pages} {3865} (\bibinfo {year} {1996})}\BibitemShut {NoStop}%
\bibitem [{\citenamefont {Bl{\"o}chl}(1994)}]{Blochl1994a}%
  \BibitemOpen
  \bibfield  {author} {\bibinfo {author} {\bibfnamefont {P.~E.}\ \bibnamefont
  {Bl{\"o}chl}},\ }\bibfield  {title} {\bibinfo {title} {Projector
  augmented-wave method},\ }\href {https://doi.org/10.1103/PhysRevB.50.17953}
  {\bibfield  {journal} {\bibinfo  {journal} {Phys. Rev. B}\ }\textbf {\bibinfo
  {volume} {50}},\ \bibinfo {pages} {17953} (\bibinfo {year}
  {1994})}\BibitemShut {NoStop}%
\bibitem [{\citenamefont {Kresse}\ and\ \citenamefont
  {Joubert}(1999)}]{Kresse1999}%
  \BibitemOpen
  \bibfield  {author} {\bibinfo {author} {\bibfnamefont {G.}~\bibnamefont
  {Kresse}}\ and\ \bibinfo {author} {\bibfnamefont {D.}~\bibnamefont
  {Joubert}},\ }\bibfield  {title} {\bibinfo {title} {From ultrasoft
  pseudopotentials to the projector augmented-wave method},\ }\href
  {https://doi.org/10.1103/PhysRevB.59.1758} {\bibfield  {journal} {\bibinfo
  {journal} {Phys. Rev. B}\ }\textbf {\bibinfo {volume} {59}},\ \bibinfo
  {pages} {1758} (\bibinfo {year} {1999})}\BibitemShut {NoStop}%
\bibitem [{\citenamefont {Peng}\ \emph {et~al.}(2021)\citenamefont {Peng},
  \citenamefont {Han}, \citenamefont {Pokharel}, \citenamefont {Li},
  \citenamefont {Hashimoto}, \citenamefont {Lu}, \citenamefont {Luo},
  \citenamefont {Guo}, \citenamefont {Wang}, \citenamefont {Cui}, \citenamefont
  {Sun}, \citenamefont {Qiao}, \citenamefont {Wilson},\ and\ \citenamefont
  {He}}]{Peng2021}%
  \BibitemOpen
  \bibfield  {author} {\bibinfo {author} {\bibfnamefont {S.}~\bibnamefont
  {Peng}}, \bibinfo {author} {\bibfnamefont {Y.}~\bibnamefont {Han}}, \bibinfo
  {author} {\bibfnamefont {G.}~\bibnamefont {Pokharel}}, \bibinfo {author}
  {\bibfnamefont {Z.}~\bibnamefont {Li}}, \bibinfo {author} {\bibfnamefont
  {M.}~\bibnamefont {Hashimoto}}, \bibinfo {author} {\bibfnamefont {D.-H.}\
  \bibnamefont {Lu}}, \bibinfo {author} {\bibfnamefont {Y.}~\bibnamefont
  {Luo}}, \bibinfo {author} {\bibfnamefont {M.}~\bibnamefont {Guo}}, \bibinfo
  {author} {\bibfnamefont {B.}~\bibnamefont {Wang}}, \bibinfo {author}
  {\bibfnamefont {S.}~\bibnamefont {Cui}}, \bibinfo {author} {\bibfnamefont
  {Z.}~\bibnamefont {Sun}}, \bibinfo {author} {\bibfnamefont {Z.-H.}\
  \bibnamefont {Qiao}}, \bibinfo {author} {\bibfnamefont {S.~D.}\ \bibnamefont
  {Wilson}},\ and\ \bibinfo {author} {\bibfnamefont {J.-F.}\ \bibnamefont
  {He}},\ }\bibfield  {title} {\bibinfo {title} {Intrinsic flat-bands, dirac
  fermions and electron-boson coupling in kagome-lattice metal
  \text{GdV$_6$Sn$_6$}},\ }\href@noop {} {\bibfield  {journal} {\bibinfo
  {journal} {arXiv preprint}\ } (\bibinfo {year} {2021})}\BibitemShut {NoStop}%
\bibitem [{\citenamefont {Anisimov}\ \emph {et~al.}(1997)\citenamefont
  {Anisimov}, \citenamefont {Aryasetiawan},\ and\ \citenamefont
  {Lichtenstein}}]{Anisimov1997}%
  \BibitemOpen
  \bibfield  {author} {\bibinfo {author} {\bibfnamefont {V.~I.}\ \bibnamefont
  {Anisimov}}, \bibinfo {author} {\bibfnamefont {F.}~\bibnamefont
  {Aryasetiawan}},\ and\ \bibinfo {author} {\bibfnamefont {A.~I.}\ \bibnamefont
  {Lichtenstein}},\ }\bibfield  {title} {\bibinfo {title} {First-principles
  calculations of the electronic structure and spectra of strongly correlated
  systems: the lda + u method},\ }\href
  {https://doi.org/10.1088/0953-8984/9/4/002} {\bibfield  {journal} {\bibinfo
  {journal} {J. Phys.: Condens. Matter}\ }\textbf {\bibinfo {volume} {9}},\
  \bibinfo {pages} {767} (\bibinfo {year} {1997})}\BibitemShut {NoStop}%
\bibitem [{\citenamefont {Mostofi}\ \emph {et~al.}(2014)\citenamefont
  {Mostofi}, \citenamefont {Yates}, \citenamefont {Pizzi}, \citenamefont {Lee},
  \citenamefont {Souza}, \citenamefont {Vanderbilt},\ and\ \citenamefont
  {Marzari}}]{Mostofi2014}%
  \BibitemOpen
  \bibfield  {author} {\bibinfo {author} {\bibfnamefont {A.~A.}\ \bibnamefont
  {Mostofi}}, \bibinfo {author} {\bibfnamefont {J.~R.}\ \bibnamefont {Yates}},
  \bibinfo {author} {\bibfnamefont {G.}~\bibnamefont {Pizzi}}, \bibinfo
  {author} {\bibfnamefont {Y.-S.}\ \bibnamefont {Lee}}, \bibinfo {author}
  {\bibfnamefont {I.}~\bibnamefont {Souza}}, \bibinfo {author} {\bibfnamefont
  {D.}~\bibnamefont {Vanderbilt}},\ and\ \bibinfo {author} {\bibfnamefont
  {N.}~\bibnamefont {Marzari}},\ }\bibfield  {title} {\bibinfo {title} {An
  updated version of wannier90: A tool for obtaining maximally-localised
  {Wannier} functions},\ }\href {https://doi.org/10.1016/j.cpc.2014.05.003}
  {\bibfield  {journal} {\bibinfo  {journal} {Comput. Phys. Commun.}\ }\textbf
  {\bibinfo {volume} {185}},\ \bibinfo {pages} {2309} (\bibinfo {year}
  {2014})}\BibitemShut {NoStop}%
\bibitem [{\citenamefont {Wu}\ \emph {et~al.}(2018)\citenamefont {Wu},
  \citenamefont {Zhang}, \citenamefont {Song}, \citenamefont {Troyer},\ and\
  \citenamefont {Soluyanov}}]{Wu2018}%
  \BibitemOpen
  \bibfield  {author} {\bibinfo {author} {\bibfnamefont {Q.-S.}\ \bibnamefont
  {Wu}}, \bibinfo {author} {\bibfnamefont {S.-N.}\ \bibnamefont {Zhang}},
  \bibinfo {author} {\bibfnamefont {H.-F.}\ \bibnamefont {Song}}, \bibinfo
  {author} {\bibfnamefont {M.}~\bibnamefont {Troyer}},\ and\ \bibinfo {author}
  {\bibfnamefont {A.~A.}\ \bibnamefont {Soluyanov}},\ }\bibfield  {title}
  {\bibinfo {title} {{WannierTools}: An open-source software package for novel
  topological materials},\ }\href {https://doi.org/10.1016/j.cpc.2017.09.033}
  {\bibfield  {journal} {\bibinfo  {journal} {Comput. Phys. Commun.}\ }\textbf
  {\bibinfo {volume} {224}},\ \bibinfo {pages} {405} (\bibinfo {year}
  {2018})}\BibitemShut {NoStop}%
\bibitem [{\citenamefont {Sancho}\ \emph {et~al.}(1985)\citenamefont {Sancho},
  \citenamefont {Sancho}, \citenamefont {Sancho},\ and\ \citenamefont
  {Rubio}}]{Sancho1985}%
  \BibitemOpen
  \bibfield  {author} {\bibinfo {author} {\bibfnamefont {M.~P.~L.}\
  \bibnamefont {Sancho}}, \bibinfo {author} {\bibfnamefont {J.~M.~L.}\
  \bibnamefont {Sancho}}, \bibinfo {author} {\bibfnamefont {J.~M.~L.}\
  \bibnamefont {Sancho}},\ and\ \bibinfo {author} {\bibfnamefont
  {J.}~\bibnamefont {Rubio}},\ }\bibfield  {title} {\bibinfo {title} {Highly
  convergent schemes for the calculation of bulk and surface {Green}
  functions},\ }\href {https://doi.org/10.1088/0305-4608/15/4/009} {\bibfield
  {journal} {\bibinfo  {journal} {J. Phys. F}\ }\textbf {\bibinfo {volume}
  {15}},\ \bibinfo {pages} {851} (\bibinfo {year} {1985})}\BibitemShut
  {NoStop}%
\bibitem [{\citenamefont {Gao}\ \emph {et~al.}(2021)\citenamefont {Gao},
  \citenamefont {Wu}, \citenamefont {Persson},\ and\ \citenamefont
  {Wang}}]{Gao2021}%
  \BibitemOpen
  \bibfield  {author} {\bibinfo {author} {\bibfnamefont {J.}~\bibnamefont
  {Gao}}, \bibinfo {author} {\bibfnamefont {Q.}~\bibnamefont {Wu}}, \bibinfo
  {author} {\bibfnamefont {C.}~\bibnamefont {Persson}},\ and\ \bibinfo {author}
  {\bibfnamefont {Z.}~\bibnamefont {Wang}},\ }\bibfield  {title} {\bibinfo
  {title} {Irvsp: To obtain irreducible representations of electronic states in
  the vasp},\ }\href
  {https://doi.org/https://doi.org/10.1016/j.cpc.2020.107760} {\bibfield
  {journal} {\bibinfo  {journal} {Comput. Phys. Commun.}\ }\textbf {\bibinfo
  {volume} {261}},\ \bibinfo {pages} {107760} (\bibinfo {year}
  {2021})}\BibitemShut {NoStop}%
\bibitem [{\citenamefont {Dronskowski}\ and\ \citenamefont
  {Bloechl}(1993)}]{Dronskowski1993}%
  \BibitemOpen
  \bibfield  {author} {\bibinfo {author} {\bibfnamefont {R.}~\bibnamefont
  {Dronskowski}}\ and\ \bibinfo {author} {\bibfnamefont {P.~E.}\ \bibnamefont
  {Bloechl}},\ }\bibfield  {title} {\bibinfo {title} {Crystal orbital
  {Hamilton} populations ({COHP}): energy-resolved visualization of chemical
  bonding in solids based on density-functional calculations},\ }\href
  {https://doi.org/10.1021/j100135a014} {\bibfield  {journal} {\bibinfo
  {journal} {J. Phys. Chem.}\ }\textbf {\bibinfo {volume} {97}},\ \bibinfo
  {pages} {8617} (\bibinfo {year} {1993})}\BibitemShut {NoStop}%
\bibitem [{\citenamefont {Deringer}\ \emph {et~al.}(2011)\citenamefont
  {Deringer}, \citenamefont {Tchougr{\'e}eff},\ and\ \citenamefont
  {Dronskowski}}]{Deringer2011}%
  \BibitemOpen
  \bibfield  {author} {\bibinfo {author} {\bibfnamefont {V.~L.}\ \bibnamefont
  {Deringer}}, \bibinfo {author} {\bibfnamefont {A.~L.}\ \bibnamefont
  {Tchougr{\'e}eff}},\ and\ \bibinfo {author} {\bibfnamefont {R.}~\bibnamefont
  {Dronskowski}},\ }\bibfield  {title} {\bibinfo {title} {Crystal orbital
  {Hamilton} population ({COHP}) analysis as projected from plane-wave basis
  sets},\ }\href {https://doi.org/10.1021/jp202489s} {\bibfield  {journal}
  {\bibinfo  {journal} {J. Phys. Chem. A}\ }\textbf {\bibinfo {volume} {115}},\
  \bibinfo {pages} {5461} (\bibinfo {year} {2011})}\BibitemShut {NoStop}%
\bibitem [{\citenamefont {Maintz}\ \emph {et~al.}(2013)\citenamefont {Maintz},
  \citenamefont {Deringer}, \citenamefont {Tchougr{\'e}eff},\ and\
  \citenamefont {Dronskowski}}]{Maintz2013}%
  \BibitemOpen
  \bibfield  {author} {\bibinfo {author} {\bibfnamefont {S.}~\bibnamefont
  {Maintz}}, \bibinfo {author} {\bibfnamefont {V.~L.}\ \bibnamefont
  {Deringer}}, \bibinfo {author} {\bibfnamefont {A.~L.}\ \bibnamefont
  {Tchougr{\'e}eff}},\ and\ \bibinfo {author} {\bibfnamefont {R.}~\bibnamefont
  {Dronskowski}},\ }\bibfield  {title} {\bibinfo {title} {Analytic projection
  from plane-wave and {PAW} wavefunctions and application to chemical-bonding
  analysis in solids},\ }\href {https://doi.org/10.1002/jcc.23424} {\bibfield
  {journal} {\bibinfo  {journal} {J. Comput. Chem.}\ }\textbf {\bibinfo
  {volume} {34}},\ \bibinfo {pages} {2557} (\bibinfo {year}
  {2013})}\BibitemShut {NoStop}%
\bibitem [{\citenamefont {Maintz}\ \emph {et~al.}(2016)\citenamefont {Maintz},
  \citenamefont {Deringer}, \citenamefont {Tchougr{\'e}eff},\ and\
  \citenamefont {Dronskowski}}]{Maintz2016}%
  \BibitemOpen
  \bibfield  {author} {\bibinfo {author} {\bibfnamefont {S.}~\bibnamefont
  {Maintz}}, \bibinfo {author} {\bibfnamefont {V.~L.}\ \bibnamefont
  {Deringer}}, \bibinfo {author} {\bibfnamefont {A.~L.}\ \bibnamefont
  {Tchougr{\'e}eff}},\ and\ \bibinfo {author} {\bibfnamefont {R.}~\bibnamefont
  {Dronskowski}},\ }\bibfield  {title} {\bibinfo {title} {{LOBSTER}: A tool to
  extract chemical bonding from plane-wave based {DFT}},\ }\href
  {https://doi.org/10.1002/jcc.24300} {\bibfield  {journal} {\bibinfo
  {journal} {J. Comput. Chem.}\ }\textbf {\bibinfo {volume} {37}},\ \bibinfo
  {pages} {1030} (\bibinfo {year} {2016})}\BibitemShut {NoStop}%
\bibitem [{\citenamefont {Momma}\ and\ \citenamefont
  {Izumi}(2011)}]{Momma2011}%
  \BibitemOpen
  \bibfield  {author} {\bibinfo {author} {\bibfnamefont {K.}~\bibnamefont
  {Momma}}\ and\ \bibinfo {author} {\bibfnamefont {F.}~\bibnamefont {Izumi}},\
  }\bibfield  {title} {\bibinfo {title} {{{\it VESTA3} for three-dimensional
  visualization of crystal, volumetric and morphology data}},\ }\href
  {https://doi.org/10.1107/S0021889811038970} {\bibfield  {journal} {\bibinfo
  {journal} {J. Appl. Crystallogr.}\ }\textbf {\bibinfo {volume} {44}},\
  \bibinfo {pages} {1272} (\bibinfo {year} {2011})}\BibitemShut {NoStop}%
\bibitem [{\citenamefont {Larsson}\ \emph {et~al.}(1996)\citenamefont
  {Larsson}, \citenamefont {Haeberlein}, \citenamefont {Lidin},\ and\
  \citenamefont {Schwarz}}]{LARSSON199679}%
  \BibitemOpen
  \bibfield  {author} {\bibinfo {author} {\bibfnamefont {A.}~\bibnamefont
  {Larsson}}, \bibinfo {author} {\bibfnamefont {M.}~\bibnamefont {Haeberlein}},
  \bibinfo {author} {\bibfnamefont {S.}~\bibnamefont {Lidin}},\ and\ \bibinfo
  {author} {\bibfnamefont {U.}~\bibnamefont {Schwarz}},\ }\bibfield  {title}
  {\bibinfo {title} {Single crystal structure refinement and high-pressure
  properties of \text{CoSn}},\ }\href
  {https://doi.org/https://doi.org/10.1016/0925-8388(95)02189-2} {\bibfield
  {journal} {\bibinfo  {journal} {J. Alloys Compd.}\ }\textbf {\bibinfo
  {volume} {240}},\ \bibinfo {pages} {79} (\bibinfo {year} {1996})}\BibitemShut
  {NoStop}%
\bibitem [{\citenamefont {Malaman}\ \emph {et~al.}(1976)\citenamefont
  {Malaman}, \citenamefont {Roques}, \citenamefont {Courtois},\ and\
  \citenamefont {Protas}}]{Malaman_1976}%
  \BibitemOpen
  \bibfield  {author} {\bibinfo {author} {\bibfnamefont {B.}~\bibnamefont
  {Malaman}}, \bibinfo {author} {\bibfnamefont {B.}~\bibnamefont {Roques}},
  \bibinfo {author} {\bibfnamefont {A.}~\bibnamefont {Courtois}},\ and\
  \bibinfo {author} {\bibfnamefont {J.}~\bibnamefont {Protas}},\ }\bibfield
  {title} {\bibinfo {title} {{Structure cristalline du stannure de fer
  \text{Fe${\sb 3}$Sn${\sb 2}$}}},\ }\href
  {https://doi.org/10.1107/S0567740876005323} {\bibfield  {journal} {\bibinfo
  {journal} {Acta Cryst. B}\ }\textbf {\bibinfo {volume} {32}},\ \bibinfo
  {pages} {1348} (\bibinfo {year} {1976})}\BibitemShut {NoStop}%
\bibitem [{\citenamefont {Lin}\ and\ \citenamefont {Chen}(2020)}]{Lin_2020}%
  \BibitemOpen
  \bibfield  {author} {\bibinfo {author} {\bibfnamefont {Z.-Z.}\ \bibnamefont
  {Lin}}\ and\ \bibinfo {author} {\bibfnamefont {X.}~\bibnamefont {Chen}},\
  }\bibfield  {title} {\bibinfo {title} {Tunable massive dirac fermions in
  ferromagnetic \text{Fe$_3$Sn$_2$} kagome lattice},\ }\href
  {https://doi.org/https://doi.org/10.1002/pssr.201900705} {\bibfield
  {journal} {\bibinfo  {journal} {Phys. Status Solidi RRL}\ }\textbf {\bibinfo
  {volume} {14}},\ \bibinfo {pages} {1900705} (\bibinfo {year}
  {2020})}\BibitemShut {NoStop}%
\bibitem [{\citenamefont {Guin}\ \emph {et~al.}(2019)\citenamefont {Guin},
  \citenamefont {Vir}, \citenamefont {Zhang}, \citenamefont {Kumar},
  \citenamefont {Watzman}, \citenamefont {Fu}, \citenamefont {Liu},
  \citenamefont {Manna}, \citenamefont {Schnelle}, \citenamefont {Gooth},
  \citenamefont {Shekhar}, \citenamefont {Sun},\ and\ \citenamefont
  {Felser}}]{Guin_2019}%
  \BibitemOpen
  \bibfield  {author} {\bibinfo {author} {\bibfnamefont {S.~N.}\ \bibnamefont
  {Guin}}, \bibinfo {author} {\bibfnamefont {P.}~\bibnamefont {Vir}}, \bibinfo
  {author} {\bibfnamefont {Y.}~\bibnamefont {Zhang}}, \bibinfo {author}
  {\bibfnamefont {N.}~\bibnamefont {Kumar}}, \bibinfo {author} {\bibfnamefont
  {S.~J.}\ \bibnamefont {Watzman}}, \bibinfo {author} {\bibfnamefont
  {C.}~\bibnamefont {Fu}}, \bibinfo {author} {\bibfnamefont {E.}~\bibnamefont
  {Liu}}, \bibinfo {author} {\bibfnamefont {K.}~\bibnamefont {Manna}}, \bibinfo
  {author} {\bibfnamefont {W.}~\bibnamefont {Schnelle}}, \bibinfo {author}
  {\bibfnamefont {J.}~\bibnamefont {Gooth}}, \bibinfo {author} {\bibfnamefont
  {C.}~\bibnamefont {Shekhar}}, \bibinfo {author} {\bibfnamefont
  {Y.}~\bibnamefont {Sun}},\ and\ \bibinfo {author} {\bibfnamefont
  {C.}~\bibnamefont {Felser}},\ }\bibfield  {title} {\bibinfo {title}
  {Zero-field nernst effect in a ferromagnetic kagome-lattice weyl-semimetal
  \text{Co$_3$Sn$_2$S$_2$}},\ }\href
  {https://doi.org/https://doi.org/10.1002/adma.201806622} {\bibfield
  {journal} {\bibinfo  {journal} {Adv. Mater.}\ }\textbf {\bibinfo {volume}
  {31}},\ \bibinfo {pages} {1806622} (\bibinfo {year} {2019})}\BibitemShut
  {NoStop}%
\bibitem [{\citenamefont {Bocarsly}\ \emph {et~al.}(2018)\citenamefont
  {Bocarsly}, \citenamefont {Need}, \citenamefont {Seshadri},\ and\
  \citenamefont {Wilson}}]{Bocarsly2018}%
  \BibitemOpen
  \bibfield  {author} {\bibinfo {author} {\bibfnamefont {J.~D.}\ \bibnamefont
  {Bocarsly}}, \bibinfo {author} {\bibfnamefont {R.~F.}\ \bibnamefont {Need}},
  \bibinfo {author} {\bibfnamefont {R.}~\bibnamefont {Seshadri}},\ and\
  \bibinfo {author} {\bibfnamefont {S.~D.}\ \bibnamefont {Wilson}},\ }\bibfield
   {title} {\bibinfo {title} {Magnetoentropic signatures of skyrmionic phase
  behavior in {FeGe}},\ }\href {https://doi.org/10.1103/PhysRevB.97.100404}
  {\bibfield  {journal} {\bibinfo  {journal} {Phys. Rev. B}\ }\textbf {\bibinfo
  {volume} {97}},\ \bibinfo {pages} {100404} (\bibinfo {year}
  {2018})}\BibitemShut {NoStop}%
\bibitem [{\citenamefont {Feng}\ \emph {et~al.}(2015)\citenamefont {Feng},
  \citenamefont {Pang}, \citenamefont {Wu}, \citenamefont {Wang}, \citenamefont
  {Weng}, \citenamefont {Li}, \citenamefont {Dai}, \citenamefont {Fang},
  \citenamefont {Shi},\ and\ \citenamefont {Lu}}]{Feng_2015}%
  \BibitemOpen
  \bibfield  {author} {\bibinfo {author} {\bibfnamefont {J.}~\bibnamefont
  {Feng}}, \bibinfo {author} {\bibfnamefont {Y.}~\bibnamefont {Pang}}, \bibinfo
  {author} {\bibfnamefont {D.}~\bibnamefont {Wu}}, \bibinfo {author}
  {\bibfnamefont {Z.}~\bibnamefont {Wang}}, \bibinfo {author} {\bibfnamefont
  {H.}~\bibnamefont {Weng}}, \bibinfo {author} {\bibfnamefont {J.}~\bibnamefont
  {Li}}, \bibinfo {author} {\bibfnamefont {X.}~\bibnamefont {Dai}}, \bibinfo
  {author} {\bibfnamefont {Z.}~\bibnamefont {Fang}}, \bibinfo {author}
  {\bibfnamefont {Y.}~\bibnamefont {Shi}},\ and\ \bibinfo {author}
  {\bibfnamefont {L.}~\bibnamefont {Lu}},\ }\bibfield  {title} {\bibinfo
  {title} {Large linear magnetoresistance in dirac semimetal
  \text{${\mathrm{Cd}}_{3}{\mathrm{As}}_{2}$} with fermi surfaces close to the
  dirac points},\ }\href {https://doi.org/10.1103/PhysRevB.92.081306}
  {\bibfield  {journal} {\bibinfo  {journal} {Phys. Rev. B}\ }\textbf {\bibinfo
  {volume} {92}},\ \bibinfo {pages} {081306} (\bibinfo {year}
  {2015})}\BibitemShut {NoStop}%
\bibitem [{\citenamefont {de~Visser}\ \emph {et~al.}(2006)\citenamefont
  {de~Visser}, \citenamefont {Ponomarenko}, \citenamefont {Galistu},
  \citenamefont {de~Lang}, \citenamefont {Pruisken}, \citenamefont {Zeitler},\
  and\ \citenamefont {Maude}}]{Visser_2006}%
  \BibitemOpen
  \bibfield  {author} {\bibinfo {author} {\bibfnamefont {A.}~\bibnamefont
  {de~Visser}}, \bibinfo {author} {\bibfnamefont {L.~A.}\ \bibnamefont
  {Ponomarenko}}, \bibinfo {author} {\bibfnamefont {G.}~\bibnamefont
  {Galistu}}, \bibinfo {author} {\bibfnamefont {D.~T.~N.}\ \bibnamefont
  {de~Lang}}, \bibinfo {author} {\bibfnamefont {A.~M.~M.}\ \bibnamefont
  {Pruisken}}, \bibinfo {author} {\bibfnamefont {U.}~\bibnamefont {Zeitler}},\
  and\ \bibinfo {author} {\bibfnamefont {D.}~\bibnamefont {Maude}},\ }\bibfield
   {title} {\bibinfo {title} {Quantum critical behaviour of the
  plateau-insulator transition in the quantum hall regime},\ }\href
  {https://doi.org/10.1088/1742-6596/51/1/088} {\bibfield  {journal} {\bibinfo
  {journal} {J. Phys.: Conf. Ser.}\ }\textbf {\bibinfo {volume} {51}},\
  \bibinfo {pages} {379} (\bibinfo {year} {2006})}\BibitemShut {NoStop}%
\bibitem [{ESI()}]{ESI}%
  \BibitemOpen
  \href@noop {} {}\bibinfo {note} {See Supplemental Information for further
  details}\BibitemShut {NoStop}%
\bibitem [{\citenamefont {Kadowaki}\ and\ \citenamefont
  {Woods}(1986)}]{KADOWAKI_1986}%
  \BibitemOpen
  \bibfield  {author} {\bibinfo {author} {\bibfnamefont {K.}~\bibnamefont
  {Kadowaki}}\ and\ \bibinfo {author} {\bibfnamefont {S.}~\bibnamefont
  {Woods}},\ }\bibfield  {title} {\bibinfo {title} {Universal relationship of
  the resistivity and specific heat in heavy-fermion compounds},\ }\href
  {https://doi.org/https://doi.org/10.1016/0038-1098(86)90785-4} {\bibfield
  {journal} {\bibinfo  {journal} {Solid State Commun.}\ }\textbf {\bibinfo
  {volume} {58}},\ \bibinfo {pages} {507} (\bibinfo {year} {1986})}\BibitemShut
  {NoStop}%
\end{thebibliography}%

\end{document}